\def\fullversionflag{0}    

\ifodd\fullversionflag
  \documentclass[runningheads]{llncs}
  \usepackage{fullpage}
\else
  \documentclass[runningheads]{llncs}
  \usepackage[T1]{fontenc}

\fi

\usepackage{etex}
\usepackage{etoolbox}
\usepackage{enumitem}

\newtoggle{fullversion}
\ifodd\fullversionflag
    \toggletrue{fullversion}
\else
    \togglefalse{fullversion}
\fi

\usepackage{float}
\usepackage{mathtools}
\usepackage{amsmath}
\usepackage{amsfonts}
\usepackage{algorithm}
\usepackage{algpseudocode}
\usepackage{tabularx}
\usepackage{graphicx}
\usepackage{listings}
\usepackage{environ}
\usepackage{enumerate}
\usepackage{xcolor}
\usepackage{bm}
\usepackage{multirow}
\usepackage{url}
\usepackage{hyperref}[colorlinks=true,citecolor=black,urlcolor=blue,linkcolor=black]
\usepackage{cleveref}
\usepackage{subcaption}

\newcommand{\ignore}[1]{}

\makeatletter
\renewcommand\section{\@startsection{section}{1}{\z@}%
  {-12\p@ \@plus -3\p@ \@minus -2\p@}%
  {8\p@ \@plus 2\p@ \@minus 1\p@}%
  {\normalfont\large\bfseries\boldmath}}
\renewcommand\subsection{\@startsection{subsection}{2}{\z@}%
  {-10\p@ \@plus -2\p@ \@minus -1\p@}%
  {6\p@ \@plus 1\p@ \@minus 1\p@}%
  {\normalfont\normalsize\bfseries\boldmath}}
\renewcommand\subsubsection{\@startsection{subsubsection}{3}{\z@}%
  {-8\p@ \@plus -2\p@ \@minus -1\p@}%
  {4\p@ \@plus 1\p@ \@minus 1\p@}%
  {\normalfont\normalsize\bfseries\boldmath}}
\makeatother

\usepackage{algpseudocode}

\algdef{SE}[FUNCTION]{Function}{EndFunction}%
   [2]{\fontsize{5.9pt}{9pt}\selectfont\algorithmicfunction\ \textproc{#1}\ifthenelse{\equal{#2}{}}{}{(#2)}}%
   {\algorithmicend\ \algorithmicfunction} %

\newcommand{\ct}{\textit{ct}}
\newcommand{\res}{\textit{res}}
\newcommand{\eid}{\textit{eid}}
\newcommand{\did}{\textit{did}}

\newcommand{\sample}[0]{\ensuremath{\xleftarrow{\$}}}

\newcommand{\cryptoSK}[0]{\ensuremath{\mathsf{sk}}}
\newcommand{\cryptoPK}[0]{\ensuremath{\mathsf{pk}}}

\newcommand{\encrypt}[0]{\ensuremath{\mathsf{Enc}}}
\newcommand{\encGen}[0]{\ensuremath{\mathsf{Enc.Gen}}}
\newcommand{\enc}[0]{\ensuremath{\mathsf{Enc.E}}}
\newcommand{\dec}[0]{\ensuremath{\mathsf{Enc.D}}}
\newcommand{\encAdd}[0]{\ensuremath{\mathsf{Enc.Add}}}
\newcommand{\encRerand}[0]{\ensuremath{\mathsf{Enc.Rerand}}}
\newcommand{\rerand}[0]{\ensuremath{\mathsf{Rerand}}}

\newcommand{\sig}[0]{\ensuremath{\mathsf{Sig}}}
\newcommand{\sigGen}[0]{\ensuremath{\mathsf{Sig.Gen}}}
\newcommand{\sigSign}[0]{\ensuremath{\mathsf{Sig.Sign}}}
\newcommand{\sigVerify}[0]{\ensuremath{\mathsf{Sig.Verify}}}

\newcommand{\hash}[0]{\ensuremath{\mathsf{H}}}
\newcommand{\hashGen}[0]{\ensuremath{\mathsf{H.Gen}}}

\newcommand{\MT}[0]
{\ensuremath{\mathsf{MT}}}
\newcommand{\MTGetRoot}[0]
{\ensuremath{\mathsf{MT.GetRoot}}}

\newcommand{\MTVerify}[0]
{\ensuremath{\mathsf{MT.Verify}}}
\newcommand{\MTGetProof}[0]
{\ensuremath{\mathsf{MT.GetProof}}}

\newcommand{\proofgen}[0]
{\ensuremath{\mathsf{zkProve}}}
\newcommand{\proofver}[0]
{\ensuremath{\mathsf{zkVerify}}}
\newcommand{\pk}[0]
{\ensuremath{\mathsf{pk}}}
\newcommand{\vk}[0]
{\ensuremath{\mathsf{vk}}}

\newcommand{\rvote}[0]{\ensuremath{\mathcal{R}_{\text{vote}}}}
\newcommand{\rdec}[0]{\ensuremath{\mathcal{R}_{\text{dec}}}}

\newcommand{\rdel}[0]{\ensuremath{\mathcal{R}_{\text{del}}}}

\newcommand{\renc}[0]{\ensuremath{\mathcal{R}_{\text{enc}}}}

\newcommand{\rencsub}[0]{\ensuremath{\mathcal{R}_{\text{encsub}}}}

\algrenewcommand{\algorithmiccomment}[1]{{\scriptsize \hfill// #1}}

\newcommand{\cvote}[0]{\ensuremath{\mathsf{vote}}}
\newcommand{\delegate}[0]{\ensuremath{\mathsf{delegate}}}
\newcommand{\undelegate}[0]{\ensuremath{\mathsf{undelegate}}}
\newcommand{\register}[0]{\ensuremath{\mathsf{register}}}
\newcommand{\unregister}[0]{\ensuremath{\mathsf{unregister}}}
\newcommand{\tally}[0]{\ensuremath{\mathsf{tally}}}
\newcommand{\esetup}[0]{\ensuremath{\mathsf{election \; setup}}}
\newcommand{\estart}[0]{\ensuremath{\mathsf{election \; start}}}
\newcommand{\setup}[0]{\ensuremath{\mathsf{setup}}}
\newcommand{\csetup}[0]{On Chain Setup}
\newcommand{\asetup}[0]{Authority Setup}
\newcommand{\creg}[0]{On Chain Delegate Registration}
\newcommand{\cunreg}[0]{On Chain Delegate Unregistration}
\newcommand{\cdelegate}[0]{On Chain Delegation}
\newcommand{\cundelegate}[0]{On Chain Undelegation}
\newcommand{\cesetup}[0]{On Chain Election Setup}
\newcommand{\cestart}[0]{On Chain Election Start}
\newcommand{\cvoting}[0]{On Chain Voting}
\newcommand{\ctally}[0]{On Chain Tally}

\newcommand{\zk}{\ensuremath{\mathsf{zk}}}

\algnewcommand\algorithmicupon{\textbf{upon receiving}}
\algnewcommand\algorithmicfrom{\textbf{from}}
\algnewcommand\upon[2]{\State\algorithmicupon\ #1 \algorithmicfrom\ #2}
\algnewcommand\algorithmicsend{\textbf{send}}
\algnewcommand\algorithmicto{\textbf{to}}
\algnewcommand\send[2]{\State\algorithmicsend\ #1 \algorithmicto\ #2}
\algnewcommand\alglog{\State \textbf{log}\;}
\algnewcommand\statetab{\State \hskip3em}
\algnewcommand\Store{\State \textbf{store}\;}
\newcommand{\lock}[0]{\ensuremath{\mathsf{lock}}}
\newcommand{\dactive}[0]{\ensuremath{\mathsf{active}}}

\newcommand{\adv}[0]{\ensuremath{\mathcal{A}}}
\newcommand{\simu}[0]{\ensuremath{\mathcal{S}}}

\newcommand{\fcontract}[0]{\ensuremath{\mathcal{F}_{\mathsf{Contract}}}}
\newcommand{\fvote}[0]{\ensuremath{\mathcal{F}_{\cvote}}}

\newcommand{\sucIdeal}[0]{\ensuremath{\mathsf{SUC-IDEAL}}}
\newcommand{\sucHybrid}[0]{\ensuremath{\mathsf{SUC-HYBRID}}}
\newcommand{\expr}[0]{\ensuremath{\mathsf{EXP}}}

\newcommand{\ZZ}{\mathbb{Z}}

\begin{document}

\title{Kite: How to Delegate Voting Power Privately}
\author{Kamilla Nazirkhanova \and Vrushank Gunjur \and X. Pilli Cruz-De Jesus \and Dan Boneh}
\institute{Stanford University\\
\email{\{nazirk, vrushank, pilli, dabo\}@stanford.edu}}

\authorrunning{Nazirkhanova et al.}

\maketitle
\begin{abstract}
Ensuring ballot secrecy is crucial to maintaining the integrity of a democratic process. Often, voting power is delegated to representatives (e.g., in congress) who subsequently vote on behalf of voters on specific issues. This delegation model is also widely used in Decentralized Autonomous Organizations (DAOs). Although several existing voting systems used in DAOs support private voting, they only offer public delegation. In this paper, we introduce Kite, a protocol that enables {\em private} delegation of voting power for DAO members, meaning that voters can freely delegate, revoke, and re-delegate their power without revealing any information about who they delegated to. Even the delegate does not learn who delegated to them. The only information that is recorded publicly is that the voter delegated or re-delegated their vote to someone. Kite accommodates both public and private voting for the delegates themselves. We analyze the security of our protocol within the Universal Composability (UC) framework. We implement Kite as an extension to the existing Governor Bravo smart contract on the Ethereum blockchain, that is widely used for DAO governance.  
Furthermore, we evaluate our implementation, showing that the most expensive operation, delegation, takes 7–167 seconds on a consumer-grade machine depending on the requested privacy level, demonstrating Kite’s practical feasibility.
\end{abstract}


\newcommand{\mypar}[1]{\medskip\noindent{\bf {#1}.}\ }
\newcommand{\func}[1]{\smallskip\noindent{\em {#1}:\ }}
\renewenvironment{quote}%
  {\list{}{\leftmargin=0.1in\rightmargin=0.1in}\item[]}%
  {\endlist}
\newcommand{\Kite}[0]{\mathsf{Kite}}
\section{Introduction}

Decentralized autonomous organizations (DAOs) consist of a loosely affiliated group of individuals who collectively oversee and manage a shared treasury.
Anyone can submit a proposal, and the DAO members vote.  If the proposal is accepted, it might be executed by the smart contract that manages the DAO. 
The proliferation of DAOs has generated renewed interest in
new voting mechanisms for DAOs~\cite{hall}, and has underscored the importance of privacy in voting.
For example, the Nouns DAO has been exploring ways to improve privacy in voting~\cite{Nounspri}.

\ignore{
The ability to participate privately in an election, without revealing one's vote, 
is essential for a well-functioning democratic process.  
A recent example is illustrated in the voting procedure of the Nouns DAO~\cite{Nounds}.  
This DAO, like many other DAOs, is using a voting system where every voter's vote is visible for everyone to see. 
Participants noticed the \href{https://discourse.nouns.wtf/t/small-grants-zk-private-voting-for-nouns-dao/3405}{following behavior}:
\begin{quote}
``Nouners many times aren't voting for what they believe is best. 
  Instead, they feel trapped in quid pro quo voting, afraid that their vote could reflect poorly on their image,
  and/or affect the likelihood of getting their own proposals through. 
  Conversely it occurs that Nouners vote in favor or against a proposal based on how the proposer voted for their past proposals.''
\end{quote}
As a result, the Nouns DAO is \href{https://discourse.nouns.wtf/t/small-grants-zk-private-voting-for-nouns-dao/3405}{looking to transition} to an end-to-end verifiable voting system where everyone can vote in private.
}

The research community has been exploring private digital voting systems
for a long time, starting with the work of Chaum~\cite{Chaum81} in 1981.
Some protocols are based on homomorphic encryption~\cite[Ch.3]{BernhardW13},
some are based on mix nets~\cite[Ch.6]{BernhardW13}, some are based on blind signatures~\cite[Ch.2]{BernhardW13}, and some are based on other mechanisms.
Modern protocols stress the notion of end-to-end verifiability, 
where the core requirement is that every voter can verify that their vote was counted as cast.~\cite{USENIX:Adida08,USENIX:BNPW24,USENIX:CCCCEHMPRSSV10}. 
We refer to~\cite{BernhardW13} for a survey of the area. 
More recently, some papers study private voting in the context of blockchains~\cite{Cicada,BCvote}.

\smallskip
In a typical voting system, private or not, every voter casts a ballot, these ballots are then tabulated,
and the final results are published.
However, this is not how voting works in DAOs. 
Since members do not have the desire or ability to vote on every proposal, 
the two most widely used governance protocols on Ethereum ---
Compound's {\em Governor Bravo}~\cite{governorBravo} 
and Open Zeppelin's {\em Governor}~\cite{governance} ---
support \emph{proxy voting}.
In proxy voting, a voter can optionally delegate their voting power to a delegate,
who votes on proposals on behalf of the voter, or delegate to themselves to vote directly.
These delegations are recorded publicly on chain.  
In addition, the delegate's voting history is also recorded publicly on chain.
The latter transparency allows a voter Alice to hold her delegate accountable for their voting record. In fact, this is not specific to DAOs and is central to representative democracy, where voters privately elect representatives who vote on their behalf publicly.
In a liquid democracy~\cite{blum2016liquid}, Alice can revoke her delegation at any time and delegate to someone else as often and as many times as she wants. Liquid democracy allows delegations to form transitive chains, with proxy voting being a special case of chains of length one.
Indeed, this logic is supported by most DAO governance contracts. 

In practice, a delegate can tell when a voter has delegated away from the delegate.
This makes voters reluctant to re-delegate because public re-delegation can cause social friction. 
Concretely, in the Nouns DAO a substantial voting power is held by delegates and plays a pivotal role in determining the outcomes of most proposals. From Nouns' inception in August 2021 to September 2025, an average of 70\% of votes were cast by delegates. This influence has consistently grown over time, with the average proposal in August 2025 seeing 78\% of votes coming from delegates. Moreover, delegates command a larger portion of the voting power, accounting for 46.8\%, in contrast to regular voters at 20.2\%, while the remainder of the voting power falls under the control of the treasury~\cite{nounsDashboard}. 

These numbers suggest that delegation is wide spread.
To improve the voter experience, DAOs governance systems would benefit from {\em private} delegation,
so that re-delegation does not incur a social cost to voters. 
Beyond private delegation one may also ask for private voting for delegates. 
However, we stress that private delegate voting is not always desirable because
it makes it harder for voters to hold their delegates to account, as it was noted in \cite{NejadgholiYC21}.

\mypar{Our work}
We design $\Kite$, a voting system that supports {\em private} delegation
for DAOs governance.
Alice can delegate her voting power to a delegate David so that no one, not even David, will know that Alice delegated to David.
Moreover, Alice can revoke her delegation at any time and re-delegate to someone else, without David's knowledge. Such design reduces social pressure or friction, so Alice is free to re-delegate her voting power without concern for how David or others might react. Privacy means that delegations are hidden by default, but the protocol does not prevent a voter from later proving how they delegated. Therefore, we do not claim secrecy, which would also require receipt-freeness.
Importantly, each voter can only delegate their own voting power, they cannot delegate someone else’s. This design ensures there are never delegation cycles, distinguishing $\Kite$ from liquid democracy systems. When the delegate David votes, $\Kite$ supports two options: either public voting, so that voters can hold David accountable for his voting record, or fully private voting for delegates.

We assume the existence of a ``computing bulletin board'' accessible to all parties. This board allows parties to post and read messages, as well as perform computations. All messages are authenticated and cannot be erased. In practice, a secure blockchain (e.g., Ethereum) provides such a bulletin board, and its transaction mechanism ensures authentication: every transaction is signed by the submitting account, so only the rightful owner of each account can submit delegation, undelegation, or voting messages.

$\Kite$ has three types of participants: voters, delegates, and a tally committee, which is technically a set of parties but, for clarity, we refer to interchangeably as the “trusted authority.” These participants interact with the voting system using the following functions:   (the detailed implementation of these functions is described in Section~\ref{sec:protocol}, and details on realizing the trusted authority as a set of parties are provided in Remark~\ref{remark:auth}). 

\def\changemargin#1#2{\list{}{\rightmargin#2\leftmargin#1}\item[]}
\let\endchangemargin=\endlist 

\begin{changemargin}{1em}{0em} 
    \func{Setup} initiated by the tally committee. 
    The tally committee produces public parameters that are used in subsequent subprotocols and sets up the on-chain contract.
    
   \func{Delegate Registration/Unregistration} called by a voter/delegate who wishes to become a delegate/stop being a delegate, respectively. It is executed on the smart contract. It takes the voter's/delegate's address as input and updates its status to 'delegate'/'voter'. 
    
    
    \func{Delegation/Undelegation} called by a voter who wants to delegate/undelegate their voting tokens to a delegate. Delegation takes the voter's and delegate's addresses. The on-chain function updates a public data structure, which reveals nothing about the delegate. Undelegation takes the voter's address and their previous delegation identifier as input. Upon execution, the on-chain function updates the public data structure, indicating that the voter has undelegated their voting tokens. Nothing else is revealed. Note that there is no need for relays as the fact that voter delegated/undelegated is public while the delegate's identity remains hidden.
    
    
    \func{Election Setup} called by a voter who wants to submit a proposal to a vote. It requires the voter's address, election ID, and a description of the election as input. Upon execution, the on-chain function returns the election parameters, which then become publicly available to all participants.
    
    \func{Election Start} called by the election creator, a voter who previously initiated the election setup. It requires the voter's address and the election ID as input. The on-chain function then returns a commitment to the token distribution as it stands at the start of the election. This is important because the token delegations might change during the election window, but the system uses the recorded delegations at the start of the election.

    \func{Voting} called by a delegate wishing to cast a vote. It requires the election ID, the delegate's address, and their vote as input. Upon execution, the on-chain function returns updated election parameters, which include an updated encrypted tally.

    \func{Tally} called by the trusted authority at the end of the election. It requires the trusted authority's secret key for the encryption scheme, the election ID and the encrypted tallies of the options as input. The function then decrypts the result and calls the on-chain function, which returns the tallies in the clear, thereby making the results public for everyone.
\end{changemargin}

\noindent
In Section~\ref{sec:security} we analyze the security of $\Kite$ using the Universal Composability (UC) framework~\cite{canetti20}. In particular, we use a variant called (SUC)~\cite{C:CanCohLin15}.
We first define an ideal voting functionality and then utilize the composition theorem to prove the security of our protocol in a hybrid settings where we rely on a provided computing bulletin board functionality, which we also formally define. In our setup, this bulletin board is implemented by a blockchain. Our formalism for the bulletin board simply abstracts the properties of the blockchain that are needed to prove security of our voting protocol.

\mypar{Implementation}
We developed a proof-of-concept in Solidity, extending the standard Governor Bravo smart contract~\cite{governorBravo} to support private delegation with public voting. Our user interface mimics Nouns DAO to demonstrate how our protocol addresses their quid pro quo voting challenges. This approach makes our implementation broadly applicable to many DAOs.

In DAO governance, the voting power of each voter is determined by the number of voting tokens they own. To manage these voting tokens, we utilize an ERC-20 contract~\cite{ERC20}, a widely-used standard for fungible tokens on the Ethereum network. 
Crucially, $\Kite$ requires a mechanism to lock delegated tokens. This operation temporarily restricts their transfer and use in order to prevent the reuse of the voting power. Without such a lock, a user could delegate their tokens and then transfer them, allowing the voting power to be used twice.

Our implementation uses zero-knowledge SNARKs~\cite{BitCanChiTro12}
and we provide implementations of all the necessary circuits.  
In Section~\ref{sec:implementation} we describe the many techniques we used to optimize proving time and reduce on-chain verification gas costs.  Our implementation uses the Noir zk-SNARK framework as the underlying ZK system. 
Finally, in Section~\ref{sec:performance} we describe the performance of the system.

%

\subsection{Related Work}
\label{sec:related}

Several works had previously studied proxy voting for general voting systems. 
To the best of our knowledge, existing work does not consider
the specific challenges and opportunities that come up in the 
context of DAO governance.

A number of works design cryptographic proxy voting systems~\cite{ZwattendorferHT13,KulykMNV16,KulykNMBV17,KulykNMV17}, addressing properties such as vote privacy, delegation privacy, coercion-resistance, and robustness. Kulyk et al.~\cite{KulykMNV16,KulykNMBV17,KulykNMV17} explored integrating proxy voting into systems like Helios~\cite{USENIX:Adida08} and boardroom voting~\cite{KulykNVFK14}, achieving delegation privacy using various tools such as anonymous channels, delegation servers, or secret-sharing. In contrast, $\Kite$ does not require additional parties or anonymous channels, nor does the delegator need to communicate with every voter, and the delegate learns nothing, not even that they were delegated to. These prior approaches are less suitable for blockchain settings, where minimizing additional infrastructure and on-chain interactions is crucial.

In another line of work~\cite{ZhangOB19,ZhangZ19,ZhangZNBO23,KovalchukZNYOR25}, delegation privacy, alongside other security guarantees, was explored in the context of blockchain governance, with the protocols rigorously proven UC-secure. These systems support delegation by allowing a voter to indicate, when casting a ballot, that their vote is delegated to a delegate, with privacy enforced via encryption and/or mix-nets. In contrast, $\Kite$ separates voting and delegation, so that voters do not need to delegate in every election. 
It also allows delegates to cast votes either publicly, enabling accountability, or privately, preserving privacy, whereas prior schemes support only private voting. Moreover, $\Kite$ achieves a much more straightforward tallying process, requiring only a single decryption of the final tally. By comparison, the schemes of~\cite{ZhangOB19,ZhangZ19,ZhangZNBO23,KovalchukZNYOR25} must process every ballot individually to compute delegations and the final result. Since these computations in practice are distributed among the tally committee, simpler operations are desirable, especially in a blockchain setting. 
Formally, while Kite also follows the simulation paradigm, our analysis is conducted in the SUC framework rather than full UC.
We also emphasize that our design naturally extends existing DAO governance protocols and can be easily adapted without any core changes.


\section{Preliminaries} 
\label{sec:prelims}

In this section we briefly list the cryptographic primitives used in our system, all of which are standard. 
\begin{itemize}
     \item A CPA-secure additively homomorphic encryption scheme $(\encGen, \enc, \dec,\\ \encAdd, \encRerand)$ 
    \item An existentially unforgeable digital signature scheme $(\sigGen, \sigSign, \sigVerify)$
    \item A collision resistant hash function $\hash$
    \item A Merkle Tree ($\MTGetRoot, \MTGetProof, \MTVerify$) from $\hash$
\end{itemize}
The complete syntax and details are provided in Appendix~\ref{app:prims}.

Additionally, we need succinct non-interactive zero-knowledge arguments.
A \textbf{Succinct Non-interactive Zero-Knowledge Argument of Knowledge} for a relation $\mathcal{R}$ consists of three algorithms ($\mathcal{G}, \proofgen, \proofver$) such that:
\begin{itemize}
    \item $\mathcal{G}(1^{\lambda}) \to \pk, \vk$: generates a prover and verification key pair.
    \item $\proofgen(\pk, R, w) \to \pi$: outputs a proof for a statement $R$ given a witness $w$.
    \item $\proofver(\vk, R, \pi) \to b \in \{0, 1\}$: verifies the validity of proof $\pi$.
\end{itemize}
We require it to be complete, succinct, and zero-knowledge (as defined in Appendix~\ref{app:zksnark}). Importantly, we need it to be knowledge-sound, however, we require a straight-line extractor \cite{C:Fischlin05}, as we further prove the security of our protocol in the universal composability framework.

Throughout our voting protocol, we make use of zero-knowledge proofs for various relations. In the subsequent sections, we omit the prover and verifier key inputs in the corresponding algorithms for brevity, and specify the relation as a subscript.

\newcommand{\TA}{\textsf{TA}}

\section{Protocol}
\label{sec:protocol}

In $\Kite$, we identify three types of entities: a set of voters $\{p_i\}_{i \in n}$, a trusted authority $\TA$, and an on-chain contract. The protocol is designed to allow voters to either vote directly or delegate their voting power. $\TA$ is trusted to tally the election results correctly and not reveal them prematurely.  We discuss practical relaxations to the trust in the $\TA$ later in this Section and in Section~\ref{sec:implementation}.

Let us begin with a high-level overview of the design of the delegation and voting protocols. The voting power of each voter is indicated by the number of governance ERC20 tokens that they own. A voter who wishes to vote on proposals must register as a delegate by calling a corresponding function at the on-chain contract. The contract operates on a secure blockchain and is responsible for tracking all participants, their status (either as voters or delegates), and the encrypted voting power of every delegate. We consider multiple-choice voting (specifically, two options in the following example) but also discuss how to adapt our protocol for ranked-choice voting.

Suppose there are only three voters, denoted $p_1, p_2,$ and $p_3$, each holding $t_1, t_2,$ and $t_3$ tokens respectively. If $p_2$ and $p_3$ are registered as delegates, the on-chain contract would store the delegate's voting power list as  $(\encrypt(0), \encrypt(t_2), \encrypt(t_3))$, using an additively homomorphic encryption scheme. When voter~$p_i$ decides to delegate their voting tokens to another voter~$p_j$, voter~$i$ creates a vector of all zeros except for the entry at index~$j$, which is set to the number of voting tokens held by $p_i$. This vector is then encrypted using the homomorphic encryption scheme and posted to the blockchain. The on-chain contract homomorphically adds this encrypted vector to the current list of delegate voting powers. 
Continuing with our example, if $p_1$ delegates to $p_3$, they would post $(\encrypt(0), \encrypt(0), \encrypt(t_1))$ to the contract. 
The contract then updates the delegates' voting power list to $(\encrypt(0), \encrypt(t_2), \encrypt(t_3+t_1))$ using the additive homomorphism. Additionally, it marks $p_1$ as a delegator and locks their tokens, preventing them from voting directly with their tokens. 

In $\Kite$, both public and private voting scenarios are supported. Let us first describe public voting. For example, if voter $p_3$ decides to vote for option~1, they can submit their vote openly. The on-chain contract then adds their encrypted total voting power, which is $\encrypt(t_3+t_1)$, into the tally for option 1 using the additive homomorphism. 

The private voting setting requires a different approach. Here, a voter must submit an encrypted vote count for each option. Specifically, when $p_3$ votes for option~1, it submits encryptions of zero for all options except the first. However, since $p_3$'s encrypted total voting power $\encrypt(t_3+t_1)$ is already accessible on the blockchain, they cannot simply send this for option~1, as it would compromise the privacy of their vote. Instead, $p_3$ re-randomizes their encrypted voting power before posting it for option~1. If there are two voting options available (yes or no), $p_3$ casts their vote for option~1 by submitting $\left(\rerand(\encrypt(t_3+t_1)), \encrypt(0)\right)$.

At the end of the election, $\TA$ decrypts the final tally for each option and publishes it on the blockchain.

Note that every voter's encryption must include a proof of correctness. Otherwise, a malicious voter could submit an ill-formed ciphertext and affect the election outcome. $\Kite$ uses two such proofs: one for correct delegation and one for correct voting, only needed in the private voting setting. 
\begin{remark}[Distributing the trusted authority]
\label{remark:auth}
For clarity, we model the trusted authority as a single party. In practice, however, it should be implemented as a set of parties, some of which may be corrupt. The tallying procedure of $\Kite$ is  simple, requiring only the decryption of a publicly available encrypted tally, and can be easily realized using an appropriate threshold decryption scheme \cite{threshold1,threshold2,threshold3}. In this setting, tally committee post their decryption shares on the chain, and once a sufficient number of shares is available, the final result can be decrypted.
\end{remark}

\begin{remark}[Ranked-choice voting]
Note that $\Kite$ can be modified to support ranked-choice voting. The public voting scenario is trivial, so we focus on private voting. Assume there are $n$ total voting options and $m$ is the number of options a voter can rank on their ballot. We maintain a tally vector with $n\cdot m$ entries, each corresponding to an option-rank pair. For example, in the case of $n = 5$ and $m = 3$, if a voter ranks options as $(1, 4, 5)$, they must submit an encryption vector where the entries corresponding to the pairs $(1,1)$, $(4,2)$, and $(5,3)$ contain the encrypted voting power, while all other entries are zero. While tallying would require decrypting individual ballots, in practice this may be unnecessary if the winner can be determined from first-choice tallies alone.
\end{remark}

\begin{remark}[Publishing approximate voting powers]
Since delegation is fully private, $\Kite$ 
might face the problem of over-delegation, where one delegate accumulates sufficient voting power to cause a proposal to pass by this single delegate's vote~\cite{strnad}. One way to resolve this issue is to periodically post a vector of approximate voting power of all delegates. More specifically, the trusted authority members decrypt the voting power vector, add differentially private noise to the vector to hide who delegated to who, and publish the noisy vector on the blockchain.  A similar issue arises in private voting systems more generally: since voters do not observe intermediate tallies, they cannot determine whether their vote is pivotal and may therefore abstain from voting. As with the delegation power vector, this can be mitigated by periodically publishing a noisy election tally but at the cost of reduced privacy.
\end{remark}

\begin{remark}[Split delegation]
As pointed out in~\cite{strnad}, it might be beneficial to allow split delegation. Although $\Kite$ supports delegation to only one delegate, it is possible to delegate to multiple delegates by splitting voting power across different accounts and delegating from each account.
\end{remark} 

\begin{remark}[Monitoring the delegate]
If Alice delegates her voting power to Bob and Bob subsequently becomes inactive or stops being a delegate, Alice’s voting power becomes unused. To avoid this, Alice should monitor Bob’s activity on the blockchain, which is publicly observable even in the private voting setting, and undelegate her voting power if necessary.
\end{remark}

In the next few subsections we walk through all the subprotocols of $\Kite$.
Along the way we describe the relations that are used in our zero-knowledge proofs to prove that all posted data is well-formed.  Glossary~\ref{tab:glossary} provided a list of all the parameters used.

\begin{table*}[t]
\makebox[\textwidth][c]{%
\setlength{\tabcolsep}{6pt}
\footnotesize
\begin{tabular}{|l|l|l|l|}
\hline
\multicolumn{4}{|c|}{\textbf{Global contract parameters}} \\ \hline
$L_{\text{\eid}}$ & List of election identifiers & $t_v$ & Number of voting tokens owned by $v$ \\ \hline
$L_{T}$ & List of token counts of eligible voters & $R_{T}$ & Merkle tree root of $L_{T}$ \\ \hline
$L_{d}$ & List of delegates' encrypted voting power & $L_{\text{\did}}$ & List of delegation identifiers \\ \hline
$R_{\text{\eid}}$ & Merkle tree root of $L_{d}$ at election start & $\mathsf{lock}$ & Lock map \\ \hline
\multicolumn{2}{|c|}{\textbf{Election parameters}} & \multicolumn{2}{c|}{\textbf{Other notation}} \\ \hline
$\eid$ & Election identifier & $\cryptoPK^{Q}_{P}, \cryptoSK^{Q}_{P}$ & Keys of party $Q$ for primitive $P$ \\ \hline
$\mathsf{vote}$ & Vote map & $\sigma_{\TA}$ & TA's signature on $R_T$ \\ \hline
$desc_{\eid}$ & Election description & & \\ \hline
$E^{\eid}_{i}$ & Encrypted votes for option $i$ & & \\ \hline
$D^{\eid}_{i}$ & Vote count for option $i$ & & \\ \hline
\end{tabular}%
}
\caption{Glossary}
\label{tab:glossary}
\end{table*}


\subsection{Setup}
In this subprotocol (illustrated in~Alg.\ref{alg:csetup}), $\TA$ begins by generating public keys for the encryption and signature schemes. The underlying plaintext space is $\ZZ_q$.  To avoid modular reduction during homomorphic addition, we set the modulus $q$ to be larger than the maximum anticipated sum of plaintext values.

The authority then creates $L_T$, list of the number of voting tokens held by each eligible voter. This list is transformed into a Merkle tree, and its root $R_T$ is signed by $\TA$. This step is crucial as participants, at various points, need to prove their token holdings. Using a Merkle tree provides a clean abstraction for the security proofs, avoiding the need to reason directly about the Ethereum state. Additionally, it reduces gas costs by eliminating on-chain calls to verify individual token balances.
$\TA$ is also tasked with updating the Merkle tree root in response to any changes in the token list. Following these steps, the trusted authority proceeds with the on-chain contract setup, executing a series of initializations. Once finished, the contract logs (writes on the blockchain) the parameters.

\begin{algorithm}[h]
    \caption{Setup}
    \label{alg:csetup}
    \begin{algorithmic}[1]
    \Function {Authority Setup}{$\lambda, L_T$} \Comment{called by the trusted authority $\TA$}
    \State $\cryptoPK^{\TA}_{\encrypt}, \cryptoSK^{\TA}_{\encrypt} \gets \encGen(1^{\lambda}, q)$ \Comment{where $q$ is twice of all the voting powers}
    \State $\cryptoPK^{\TA}_{\sig}, \cryptoSK^{\TA}_{\sig} \gets \sigGen(1^{\lambda})$
    \State $R_T \gets \MTGetRoot(L_T)$
    \State $\sigma_{T}  \gets \sigSign(\cryptoSK^{\TA}_{\sig}, R_T)$
    \State $\pk_{\zk}, \vk_{\zk} \gets \mathcal{G}(1^{\lambda}) $
    \State $L_T, R_T, L_{\text{\eid}}, \bm{L_d}, L_{\did}, \lock, \dactive \gets \Call{\csetup}{\cryptoPK^{\TA}_{\encrypt}$, $\cryptoPK^{\TA}_{\sig}$, $L_T$, $R_T$, $\sigma_{\TA}, \pk_{\zk}, \vk_{\zk}}$
    \Store $\cryptoSK^{\TA}_{\encrypt}, \cryptoSK^{\TA}_{\sig}$
    \EndFunction
    \Function{\csetup}{$ \cryptoPK^{\TA}_{\encrypt}$, $\cryptoPK^{\TA}_{\sig}$, $L_T$ $R_T$, $\sigma_{\TA}, \pk_{\zk}, \vk_{\zk}$}
    \State $L_{\text{\eid}}\gets \emptyset$ \Comment{list of election identifiers}
    \State $L_{\did} \gets \emptyset$ \Comment{list of delegation identifiers}
    \State $\lock  \gets \bm{0}$ \Comment{lock map}
    \State $\dactive  \gets \bm{0}$ \Comment{active delegate map}
    \State $\bm{L_d} \gets \bm{0}$ \Comment{list of delegates voting power, all-zero vector}
    \If{$\sigVerify(\cryptoPK^{\TA}_{\sig}, R_T, \sigma_T) = 1$} 
    \State $R_T \gets R_T$ \Comment{save $R_T$ in the contract state}
    \alglog $L_T, R_T, L_{\text{\eid}}, \bm{L_d}, L_{\did}, \lock, \dactive, \pk_{\zk}, \vk_{\zk}$
    \Else
    \State abort
    \EndIf
    
    \EndFunction
    \Function{Voter Setup}{$L_T$} \Comment{called by $p_i$}
    \State $t_{p_i} \gets L_T[p_i]$ \Comment{store the number of tokens $p_i$ has}
    \State $\bm{ct} \gets \bm{0}$ \Comment{initialize delegation vector}
    \Store $t_{p_i}, \bm{ct}$
    \EndFunction
    \end{algorithmic}
\end{algorithm}

\subsection{Delegate Registration and Unregistration}

When a voter decides to become a delegate, they call \textsc{\creg} (Alg.~\ref{alg:dreg}). This triggers the on-chain contract to mark the voter as an active delegate, simultaneously locking their funds. Again, the locking step ensures the voter cannot reuse the same tokens in the voting process. We denote the function for locking tokens by $\lock$, so calling $\lock(p_i)$ locks the tokens held by voter $p_i$. The corresponding function $\mathsf{unlock}$ reverses this operation.
Additionally, the contract updates the list of delegates' encrypted voting power. It does this by adding the encryption of the voter's power into the corresponding cell in the list $L_{d}$. 

To unregister as a delegate, the protocol follows a reverse process and calls \textsc{\cunreg} (Alg.~\ref{alg:dreg}). The on-chain contract marks the delegate as inactive and unlocks their funds.
Furthermore, the contract updates the list of delegates' encrypted voting power by subtracting the encryption of the delegate's voting power from the corresponding cell of $L_{d}$.

There are four possible state combinations a voter can encounter, based on their active and token-locking status: unlocked and inactive (has not delegated), unlocked and active (impossible), locked and inactive (delegated tokens), or locked and active (delegate). 

\begin{algorithm}[h]
    \caption{Delegate Registration/Unregistration}
    \label{alg:dreg}
    \begin{algorithmic}[1]
    \Function{Delegate Registration}{$p_i$} \Comment{called by $p_i$}
    \State $p_i, \lock[p_i], \dactive[p_i], L_d[p_i] \gets \Call{\creg}{p_i}$
    \EndFunction
    \Function{\creg}{$p_i$} 
    \If{$\dactive[p_i] = 0 \land \lock[p_i] = 0 $}
    \State $\lock(p_i)$ \Comment{lock tokens of msg.sender in ERC20 contract}
    \State $\lock[p_i] = 1 $ \Comment{update lock map}
    \State $\dactive[p_i] = 1 $ \Comment{update active delegate map}
    \State $t_{p_i} \gets L_T[p_i]$
    \State $L_d[p_i] \gets  \enc(\cryptoPK^{\TA}_{\encrypt}, t_{p_i}; 0)$ \Comment{update $p_i$'s voting power in the delegate list}
    \EndIf
    \alglog $p_i, \lock[p_i], \dactive[p_i], L_d[p_i]$
    \EndFunction
    \Function{Delegate Unregistration}{$p_i$}
    \Comment{called by $p_i$}
    \State $p_i, \lock[p_i], \dactive[p_i], L_d[p_i] \gets \Call{\cunreg}{p_i}$
    \EndFunction
    \Function{\cunreg}{$p_i$}
    \If{$\dactive[p_i] = 1 \land \lock[p_i] = 1$}
    \State $\mathsf{unlock}(p_i)$ \Comment{unlock tokens of msg.sender in ERC20 contract}
    \State $\lock[p_i] = 0 $ \Comment{update lock map}
    \State $\dactive[p_i] = 0 $ \Comment{update active delegate map}
    \State $t_{p_i} \gets L_T[p_i]$
    \State $e \gets \enc(\cryptoPK^{\TA}_{\encrypt}, -t_{p_i}; 0) $
    \State $L_d[p_i] \gets  \encAdd( \cryptoPK^{\TA}_{\encrypt}, L_d[p_i], e)$ 
    \EndIf
    \alglog $p_i, \lock[p_i], \dactive[p_i], L_d[p_i]$
    \EndFunction
    \end{algorithmic}
\end{algorithm}


\subsection{Delegation and Undelegation}
In the delegation subprotocol, Alg.~\ref{alg:delegation}, if a voter $p_i$ wishes to delegate their voting power $t_{p_i}$ to a delegate $p_j$, they create an encrypted vector $\bm{ct}$. It is an encryption of the all-zero vector, except for the entry corresponding to $p_j$. Additionally, the voter generates a proof that $\bm{ct}$ is well-formed and that they indeed possess $t_{p_i}$ tokens. This involves a Merkle inclusion proof in the tree with the root $R_T$, which was earlier uploaded by $\TA$. 

The on-chain contract then verifies that the voter's tokens are not locked and checks the proof of correctness. If these checks are passed, it locks the tokens and performs a homomorphic addition of $\bm{ct}$ to the existing list of encrypted powers of the delegates. This addition ensures that the total power of $p_j$ is accurately updated to include $t_{p_i}$.
Additionally, it computes a commitment to $\bm{ct}$, using a hash function. This hash, or commitment, is then added to the list of delegation identifiers. This step is crucial for the undelegation process that may follow later. 

During the undelegation process, Alg.~\ref{alg:delegation}, a voter essentially reverses the actions taken in the delegation phase. This includes unlocking their tokens and deducting their contributed voting power from the delegate's total. The key to executing the second part correctly lies in the use of delegation identifiers, established earlier.

For undelegation, the voter is required to present the original  delegation vector, $\bm{ct}$, that they used for delegation. Note that making the voter resubmit $\bm{ct}$ avoids storing the full ciphertext in the contract. The contract then verifies whether the hash of this is stored in $L_{\did}$. Finding the hash indicates the voter had indeed delegated their tokens using this specific $\bm{ct}$. Once confirmed, the contract proceeds to homomorphically subtract $\bm{ct}$ from $L_d$ and also removes the corresponding delegation identifier $\hash(\bm{ct})$ from $L_{\did}$.

The relation for the corresponding zero-knowledge proof is defined as follows:

\begin{multline}
\label{eq:rdel}
\rdel := \bigg\{(p_j, \bm{r}), (\cryptoPK^{\TA}_{\enc}, \bm{ct}, t_{p_i}, R_T, \pi_{p_i}, p_i)
\bigg| 
\MTVerify(t_{p_i}, p_i, \pi_{p_i}, R_T) = 1 \land \\
\qquad ct[p_j] = \enc(\cryptoPK^{\TA}_{\encrypt}, t_{p_i}; r_{j}) \land 
\forall i \neq p_j : ct[i] = \enc(\cryptoPK^{\TA}_{\encrypt}, 0; r_i)
\bigg\}
\end{multline}


In the subsequent Table~\ref{tab:rdel}, we delineate the witness and public parameters.
Note that the corresponding circuit is linear in the number of delegates. One possible optimization is to select a random subset of delegates of smaller size, which we call the \emph{anonymity set}, and construct $\bm{ct}$ only for that set. We discuss this in more detail in Section~\ref{sec:implementation} and Appendix~\ref{app:implementation}.
\begin{table*}[t]
\makebox[\textwidth][c]{%
\setlength{\tabcolsep}{6pt}
\footnotesize
\begin{tabular}{|l|l|l|l|}
\hline
\multicolumn{4}{|c|}{\textbf{Witness and Public Statement for $\rdel$}} \\ \hline
\multicolumn{2}{|c|}{\textbf{Witness}} & \multicolumn{2}{c|}{\textbf{Public Statement}} \\ \hline
$p_j \in [m]$ & Delegate's address and index & $\cryptoPK^{\TA}_{\encrypt} \in \mathbb{G}$ & Encryption public key of TA \\ \hline
$\bm{r} \in \mathbb{Z}_q^m$ & Randomness vector for encryption & $\bm{ct} \in \mathbb{G}^m$ & Delegation vector of encryptions \\ \hline
& & $t_{p_i} \in \mathbb{Z}_q$ & Voting power of voter $p_i$ \\ \hline
& & $R_T \in \mathbb{F}_p$ & Root of Merkle tree of voting powers \\ \hline
& & $\pi_{p_i} \in \mathbb{F}_p^{\log n}$ & Merkle proof for element at index $p_i$ \\ \hline
& & $p_i \in [n]$ & Voter's address and index \\ \hline
\end{tabular}%
}
\caption{Witness and Public Statement for $\rdel$}
\label{tab:rdel}
\end{table*}
\begin{algorithm}[h]
    \caption{Delegation/Undelegation}
    \label{alg:delegation}
    \begin{algorithmic}[1]
    \Function{Delegation}{$p_i, p_j$}
    \Comment{called by $p_i$} 
    \State $\bm{r} \sample \mathbb{Z}_q^m$ \Comment{$m$ is the number of registered delegates}
    \For{$i \in len(L_d) \land i \neq p_j$}
    \State $\ct[i] \gets \enc(\cryptoPK^{\TA}_{\encrypt}, 0; r_i)$
    \EndFor
    \State $\ct[p_j] \gets \enc(\cryptoPK^{\TA}_{\encrypt}, t_{p_i}; r_j)$
    \State $\pi_{p_i} \gets \MTGetProof(L_{T}, p_i)$ \Comment{get a Merkle proof for a leaf with index $p_i$}
    \State $\pi \gets \proofgen_{\rdel}\left((\cryptoPK^{\TA}_{\enc}, \bm{ct}, t_{p_i}, R_T, \pi_{p_i}, p_i), (p_j, \bm{r})\right)$
    \State $p_i, \lock[p_i], L_d, L_{\did}[p_i] \gets \Call{\cdelegate}{p_i, \bm{ct}, \pi, \pi_{p_i}}$
    \EndFunction
    \Function{\cdelegate}{$p_i, \bm{ct}, \pi, \pi_{p_i}$}
    \If {$\lock[p_i] = 0$}   \Comment{verify if $p_i$'s tokens are not locked} 
    \If{$\proofver_{\rdel}\left((\cryptoPK^{\TA}_{\enc}, \bm{ct}, t_{p_i}, R_T, \pi_{p_i}, v), \pi\right) = 1$}
    \State $\lock(p_i)$ \Comment{lock tokens of msg.sender in ERC20 contract}
     \State $\lock[p_i] = 1 $ \Comment{update lock map}
    \State $L_d \gets \encAdd( \cryptoPK^{\TA}_{\encrypt}, L_d, \bm{ct})$  \Comment{homomorphic addition of encrypted vectors}
    \State $L_{\did}[p_i] \gets \hash(\bm{ct})$ \Comment{update the list of delegation identifiers}
    \EndIf
    \EndIf
    \alglog $p_i, \lock[p_i], L_d, L_{\did}[p_i]$
    \EndFunction
        \Function{Undelegation}{$p_i, \bm{ct}$}
    \Comment{called by $p_i$} 
    \State $p_i, \lock[p_i], L_d, L_{\did}[p_i] \gets \Call{\cundelegate}{p_i, \bm{ct}}$
    \EndFunction
    \Function{\cundelegate}{$p_i, \bm{ct}$}
   
    \If {$\lock[p_i] = 1$}   
    \If{$L_{\did}[p_i] = \hash(\bm{ct})$}
    \State $\mathsf{unlock}(p_i)$ \Comment{unlock tokens of msg.sender in ERC20 contract}
    \State $\lock[p_i] = 0$ \Comment{update lock map}
    \State $L_d \gets \encAdd( \cryptoPK^{\TA}_{\encrypt}, L_d , - \bm{ct})$ \Comment{homomorphic subtraction of encrypted vectors}
    \State $L_{\did}[p_i] \gets 0$ \Comment{update the list of delegation identifiers}
    \EndIf
    \EndIf
    \alglog $p_i, \lock[p_i], L_d, L_{\did}[p_i]$
    \EndFunction
\end{algorithmic}
\end{algorithm}

   

\begin{remark}[Topic-based delegation] While we focus on a single global delegation relation, the protocol can be extended to support topic-based delegation. In this variant, each topic is associated with a separate delegation vector. For example, Alice could delegate her voting power to Bob for proposals related to the US and to Charlie for proposals related to Europe by submitting two separate encrypted delegation vectors, one for each topic-delegate pair.
\end{remark}

\subsection{Election Setup and Election Start}
Our election subprotocol is structured into two phases. The first phase begins with the election setup, initiated when the election details are made available on-chain. The second phase begins with the initiation of the voting process. Both phases are illustrated in~Alg.~\ref{alg:esetup}.
Any voter with the intention to create an election can trigger the setup phase. They are required to provide a unique election identifier and a description of the election. The contract, in response, initializes counters for each election option. In our example, we have three options: 'Yes', 'No', and 'Abstain', with their counters initialized to the encryption of zeros. Additionally, the contract establishes a vote map to track participants voting activity, preventing double voting. It also initializes a snapshot of the voting power, which is crucial for the next phase.
This phase can only be initiated by the creator of the election. At the start of each election, the contract captures a snapshot of the current root of the voting power list -- $R_{\eid}$. This snapshot is necessary because voting power can fluctuate if voters transfer tokens. By capturing a snapshot of the voting power distribution at the beginning of each election, we maintain a consistent reference point for the duration of that election. This design is particularly important when multiple elections run concurrently: each election relies on its own snapshot, preventing interference between overlapping elections with different start times. As a result, voters may freely transfer tokens without affecting the correctness of any ongoing tally, and the protocol avoids the unrealistic requirement of freezing token transfers during elections.
\begin{algorithm}[h]
    \caption{Election Setup and Start}
    \label{alg:esetup}
    \begin{algorithmic}[1]
    \Function{Election Setup}{$p_i, \eid, desc$} \Comment{called by $p_i$}
    \State $L_{\text{\eid}}, \bm{E}^{\eid}, \mathsf{vote}_{\eid}, R_{\eid}, c_{\eid} \gets \Call{\cesetup}{p_i, \eid, desc}$
    \EndFunction
    \Function{\cesetup}{$v, \eid, desc$}
    \If{$\eid \notin  L_{\text{\eid}}$}
    \State $L_{\text{\eid}} \gets L_{\text{\eid}} \cup \eid$ \Comment{update the list of election identifiers}
    \State $desc_{\eid} \gets desc$
    \State $\bm{ct_0} \gets \enc(\cryptoPK^{\TA}_{\encrypt}, \bm{0}; \bm{0})$  
    \State $\bm{E}^{\eid} \gets \bm{ct_0}$ \Comment{initialize the voting counters}
    \State $\mathsf{vote}_{\eid}  \gets \emptyset$ \Comment{initialize the election voting map}
    \State $R_{\eid} \gets 0$ 
    \Comment{initialize the voting power snapshot}
    \State $c_{\eid} \gets p_i$ \Comment{save the election creator}
     \alglog $L_{\text{\eid}}, \bm{E}^{\eid}, \mathsf{vote}_{\eid}, R_{\eid}, c_{\eid}$
     \Else
     \State abort
    \EndIf
    \EndFunction
    \Function{Creator Election Start}{$p_i, \eid$} \Comment{called by $p_i$}
    \State $R_{\eid} \gets \Call{\cestart}{p_i, \eid}$
    \State $L^{\eid}_d \gets L_d$ 
    \State \Return $L^{\eid}_d$
    \EndFunction
    \Function{\cestart}{$p_i, \eid$} 
    \If{$c_{\eid} = p_i$} \Comment{check if $p_i$ is the election creator}
    \State $R_{\eid} \gets \MTGetRoot(L_d)$ \Comment{the voting power snapshot}
    \EndIf
    \alglog $R_{\eid}$
    \EndFunction
    \end{algorithmic}
\end{algorithm}


\subsection{Voting}
For clarity, we assume there are three vote options -- Yes, No, and Abstain. However, any number of choices can be supported.
In the public voting scenario (Alg.~\ref{alg:voting}), a delegate who wants to cast their vote must send it directly to the on-chain contract. 
Along with their vote, the delegate provides a Merkle proof for their encrypted voting power, allowing the on-chain contract to verify it against the Merkle tree root $R_{\eid}$, generated at the start of the election.
The on-chain contract then performs several checks: it verifies that the delegate is active, confirms that they haven't voted previously, and validates the Merkle proof. Only after successfully passing these checks does the contract homomorphically add the delegate's encrypted voting power to the tally for the selected option.

While our protocol supports both private and public voting, delegation is kept private in all settings. The detailed algorithm for private voting is omitted here for brevity. Appendix~\ref{app:private_voting} contains the full description. The remainder of this section focuses on public voting, which is the setting for our security analysis.
\begin{remark}[Extension: Changing votes]
We also note that $\Kite$ can be extended to support vote changes. In this case, whenever a voter submits a new vote, the contract subtracts the voter’s previous contribution from the tally before adding the new one either directly in cleartext for public voting, or by homomorphically subtracting the previously submitted encrypted voting vector in the private voting setting.
\end{remark}

\begin{algorithm}[h]
    \caption{Public Voting}
    \label{alg:voting}
    \begin{algorithmic}[1]
    \Function{Voting}{$\eid, L^{\eid}, p_i, v$} \Comment{called by $p_i$}
    \If{$v \in \{\text{yes}, \text{no}, \text{abstain}\}$}
    \State $\pi_{p_i} \gets \MTGetProof(L^{\eid}_d, p_i)$
    \State $\mathsf{vote}[p_i], E^{\eid}_{v} \gets \Call{\cvoting}{\eid, d, v, \pi_{p_i}, L^{\eid}_d[p_i]}$
    \Else
    \State abort
    \EndIf
    \EndFunction 
    \Function{\cvoting}{$\eid, p_i, v, \pi_{p_i}, L^{\eid}_d[p_i]$}
    \If {$R_{\eid} \neq 0 \land \dactive[p_i] = 1 \land \mathsf{vote}[p_i] = 0  \land \MTVerify(L^{\eid}_d[p_i], p_i, \pi_{p_i}, R_{\eid}) = 1$ }   
    \State $\mathsf{vote}[p_i] = 1$
    \State $E^{\eid}_{v} \gets \encAdd( \cryptoPK^{\TA}_{\encrypt}, E^{\eid}_{v}, L^{\eid}_d[p_i])$ \Comment{homomorphic addition of the vote to the tally}
    \alglog $\mathsf{vote}[p_i], E^{\eid}_{v}$
    \Else
    \State abort
    \EndIf
    \EndFunction
    \end{algorithmic}
\end{algorithm}


\subsection{Tally}
In the Tally subprotocol, Alg.~\ref{alg:result}, $\TA$ first decrypts the tallies for each option. Subsequently, it computes the vote percentages for each option.
The final results only reflect the proportional distribution of votes, without disclosing the absolute voting power behind each option.  This is important in scenarios where a number of participating delegates is small.
If only a small number of delegates vote, the final result might reveal their total voting powers in the public vote setting. To address this, we can change $\TA$ and make it reveal the winning option only.



\begin{algorithm}[h]
    \caption{Tally}
    \label{alg:result}
    \begin{algorithmic}[1]
    \Function{Authority Tally}{$\cryptoSK^{\TA}_{\encrypt}, \eid, \bm{E}^{\eid}$}
    \Comment{called by $\TA$}
    \State $\bm{D}^{\eid} \gets \dec(\cryptoSK^{\TA}_{\encrypt}, \bm{E}^{\eid})$
    \For {$i \in \{\text{yes}, \text{no}, \text{abstain}\}$}
    \State $\res^{\eid}_i \gets 100\frac{D^{\eid}_i}{\sum^3_{i=1} D^{\eid}_i}$
    \EndFor
    \State $\res_{\eid} \gets \Call{\ctally}{\eid, \res^{\eid}}$ 
    \EndFunction
    \Function{\ctally}{$\eid, \res^{\eid}$}
    \alglog $\res^{\eid}$
    \EndFunction
    \end{algorithmic}
\end{algorithm}

\section{Security}
\label{sec:security}

\subsection{The Universal Composability Framework}

We analyze $\Kite$ using a variant of the Universal Composability (UC) framework called SUC~\cite{C:CanCohLin15}, which simplifies UC by assuming a fixed set of parties and built-in authenticated channels. Security is defined by comparing executions in the real, ideal, and hybrid models: informally, a protocol is secure if no efficient environment can distinguish between interacting with the real protocol and with an ideal functionality. Subroutines are executed locally rather than as separate interactive Turing machines, which simplifies composition. These changes make SUC easier to apply while still supporting security proofs that can be translated to standard UC. In the interest of space, and since UC is a well-established model in cryptography, we defer a more detailed description and intuition around it to Appendix~\ref{app:security}.

\subsection{Security Proof}
We prove that $\Kite$ securely realizes an ideal functionality, $\fvote$. In contrast to the UC functionalities of prior work~\cite{ZhangOB19,ZhangZ19,ZhangZNBO23,KovalchukZNYOR25}, our $\fvote$ is an SUC functionality that separates delegation from voting, enabling delegations to be long-lived across multiple elections. The proof is conducted in the $\fcontract$-hybrid model, where $\fcontract$ abstracts the bulletin board functionality provided by a smart contract. We assume a static adversary that may corrupt voters but not the trusted authority~$\TA$, and we work in the local random oracle model, which allows the simulator to program the random oracle~\cite{grom}. Note that we can also use a global random oracle but with domain separation: each party and protocol instance queries the random oracle with a unique prefix.
We note that our proof can be similarly adapted for the private voting scenario.

\begin{theorem}
$\Kite$ (Sec.~\ref{sec:protocol})
SUC-securely realizes $\fvote$ with respect to $\Pi^{\fcontract}$, 
assuming a collision-resistant hash function, 
a CPA-secure encryption scheme, and 
a secure non-interactive zero-knowledge argument of knowledge.
\end{theorem}

The full proof, including the formal functionalities $\fvote$ and $\fcontract$, as well as the simulator code and the simulation analysis, is provided in Appendix~\ref{app:security}.

\section{Implementation}
\label{sec:implementation}
We developed a proof-of-concept implementation of $\Kite$, supporting private delegation and public voting~\footnote{\url{https://anonymous.4open.science/r/GovernorPrivate-BCD5/README.md}}. The implementation consists of a React frontend, two Rust servers for cryptographic operations and trusted authority functions, Noir circuits for zero-knowledge proofs, and on-chain Ethereum smart contracts extending OpenZeppelin ERC20 and Compound Governor Bravo. The Rust backend handles proof generation, ElGamal encryption, and homomorphic operations over ElGamal ciphertexts, while the trusted authority server decrypts final tallies and posts zk-proofs of correct decryption. 

To improve efficiency and preserve privacy, $\Kite$ uses an \emph{anonymity set} during delegation. When a voter delegates, the backend constructs a set of $T$ delegate accounts that includes the intended delegate and $T-1$ randomly sampled other delegates. The voter posts encrypted voting power for all accounts in this set, along with a zero-knowledge proof of correctness. This ensures that observers know the voter delegated to someone in the set, but cannot identify the specific delegate. The size of the anonymity set balances privacy and efficiency: larger sets provide stronger privacy but increase proof size and on-chain costs, while smaller sets reduce costs at the expense of some privacy. Choosing $T=1$ reduces to public delegation.
This “hide within a set” approach has been used in prior voting protocols. For instance, in Selections\cite{ClarkH11}, a voter’s real password is hidden among a large set of panic passwords.
In contrast to that setting, in Kite the size of the set primarily affects the adversary’s success probability in guessing which delegate was chosen, rather than enabling a concrete attack. Even if the set is largely composed of corrupt delegates, the adversary cannot learn the specific recipient. It is also important to note that delegations are expected to be long-lived: a voter can keep an existing delegation to the same delegate across many elections, eliminating the need to resubmit fresh anonymity sets and thereby mitigating set-intersection linkability attacks.

All zero-knowledge relations are verified on-chain using smart contracts auto-generated from Noir, with custom optimizations to reduce proving and gas costs. For more details such as circuit design, smart contract logic, and ZK relation specifications, we refer the reader to Appendix~\ref{app:implementation}.

\section{Evaluation}
\label{sec:performance}
We evaluated the performance of our proof-of-concept implementation of $\Kite$, focusing on zero-knowledge proof generation and verification. Proving times increase with the size of the anonymity set used for delegation: smaller sets provide faster proofs but less privacy, while larger sets increase proof time and memory. For typical anonymity sets of size 5, 10, or 20, proving times remain practical, ranging from a few seconds to just over a minute.

Other ZK relations, such as vote proofs and implementation-specific operations, are lightweight in comparison. On-chain verification costs are moderate, and combining multiple relations into a single proof reduces gas usage. Using a master verifier contract significantly lowers deployment costs compared to deploying individual verifier contracts. Overall, the system demonstrates reasonable performance on consumer-grade hardware. We also note that delegation, though more costly to prove, is infrequent, so typical user interactions remain low-latency.

\label{app:graph}
\begin{figure*}[h]
    \makebox[\textwidth][c]{%
        \includegraphics[scale=0.32]{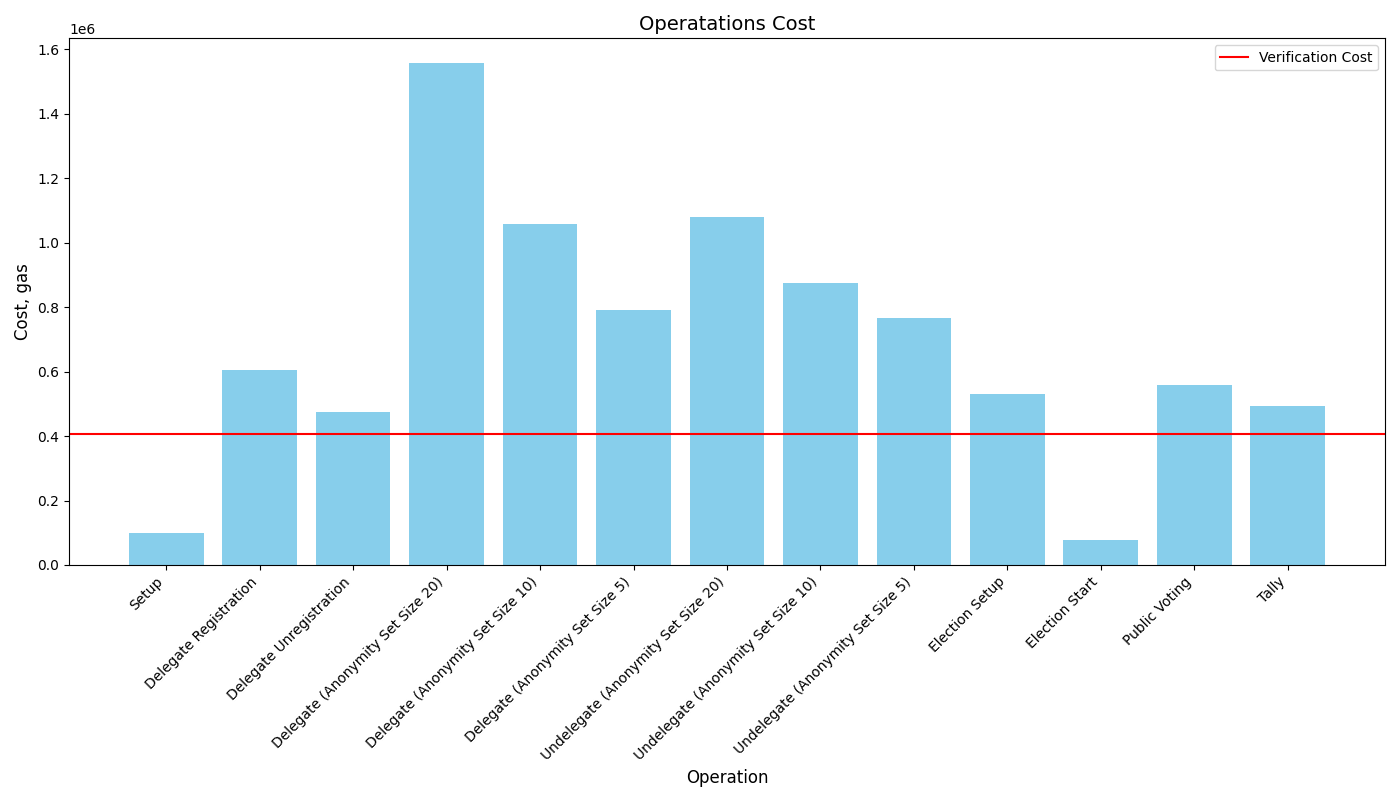}%
    }
    \caption{Gas cost of all protocol operations.}
    \label{fig:gas}
\end{figure*}
The end-to-end implementation was deployed to a local Anvil testnet. We measured the gas cost of each operation required for end-to-end voting and delegation in our proof-of-concept implementation(Fig.~\ref{fig:gas}). The gas cost associated with PLONK proof verification on chain is independent of the relation being verified, the average associated gas cost is represented by the red horizontal line. As expected, delegation and undelegation are the most resource-intensive operations due to the need for verifying homomorphic vector operations. In turn, delegate unregistration has the lowest gas cost due to its minimal inputs and on-chain operations. 
For a small number of options (here, three), private voting would not incur significant additional gas overhead: its cost is comparable to delegation with an anonymity set of size three, and thus remains on the same order as public voting.

More results, including detailed proving times, circuit sizes, gas costs, and additional graphs are provided in Appendix~\ref{app:evaluation}.

\section{Conclusion and Future Work}
\label{sec:concl}

We presented $\Kite$, a voting system for DAOs, that enables {\em private} delegation of voting power.  
The system is implemented as a direct extension to a popular DAO voting smart contract.  
Currently, $\Kite$ provides either complete transparency for delegate votes (so that delegates can be held accountable for their voting record) or total privacy for delegate votes, where only the delegate knows their own votes.  
One direction for future research is something in between, namely a system that maintains confidentiality of the delegate's vote from the general public, but reveals how they voted to voters who have delegated their voting tokens to that delegate. Another important open direction is to achieve a stronger property of ballot secrecy which would require the integration of a coercion-resistant voting mechanism in $\Kite$. Existing techniques, such as fake credentials, rely on hiding voters’ true public keys and therefore do not directly apply to DAOs, where voting power is tied to public keys.


\ignore{
In the representative democracy model, individuals can delegate their votes to delegates, who then vote on their behalf. While delegation in traditional election systems is typically private, this is not the case with electronic voting systems used in DAOs.

In this paper, we address this issue by presenting a protocol that enables private delegation of voting power. Our system supports public and private voting mechanisms. We provide a proof-of-concept implementation of the protocol evaluated on a consumer-grade device, and observed that all associated operations ran with reasonable time and memory costs.

To conclude, another practical direction is to look at reducing on-chain verification gas costs. 
One option is to outsource much of the on-chain Solidity contract to an off-chain ZK prover,
as was recently proposed in the Bonsai Governance showcase~\cite{bonsai}. 
}

{\bf Acknowledgments.} This work was funded by IOG, NSF, DARPA,
the Simons Foundation, UBRI, and NTT Research. Opinions, findings, and conclusions or recommendations expressed in this material are those of the authors and do not necessarily reflect the views
of DARPA.
\bibliographystyle{abbrv}
\bibliography{main,abbrev2,crypto}

\appendix
\section{Cryptographic Primitives}
\label{app:prims}
\subsection{Additively Homomorphic Encryption}
For our purposes, we require a additively homomorhpic \textbf{asymmetric encryption scheme} $\encrypt$ comprised of five algorithms $(\encGen, \enc, \dec, \encAdd, \encRerand)$ such that:
\begin{itemize}
    \item $\encGen(1^\lambda, q) \to (\cryptoPK, \cryptoSK)$ : generates a public and private key pair.
    \item $\enc(\cryptoPK, m ; r) \to \ct$: encrypts a message $m \in \ZZ_q$ with public key $\cryptoPK$ with randomness $r$.
    \item $\dec(\cryptoSK, \ct) \to m \in \ZZ_q$: decrypts ciphertext $ct$ using secret key $\cryptoSK$.
    \item $\encAdd(\cryptoPK, \ct_1, \ct_2) \to \ct_{+}$: homomorphically adds ciphertexts $\ct_1$ and $\ct_2$ such that $\dec(\cryptoSK, \ct_1) + \dec(\cryptoSK, \ct_2) = \dec(\cryptoSK, \ct_{+})$.
    \item $\encRerand(\cryptoPK, \ct, r') \to \ct'$: re-randomizes $\ct$ with new randomness $r'$. 
\end{itemize}
We only require that the scheme be semantically secure against chosen plaintext attack, also known as CPA-secure~\cite[Ch. 5]{BonehShoup}. 

\subsection{Digital Signature Scheme}

A \textbf{digital signature scheme} $\sig$ is a triple of algorithms $(\sigGen, \sigSign, \sigVerify)$ such that:
\begin{itemize}
    \item $\sigGen(1^\lambda) \to (\cryptoPK, \cryptoSK)$ : generates a public and private key pair.
    \item $\sigSign(\cryptoSK, m) \to \sigma$: signs message $m$ with secret key $\cryptoSK$.
    \item $\sigVerify(\cryptoPK, m, \sigma) \to b \in \{0, 1\}$:  verifies signature $\sigma$.
\end{itemize}
We require that the scheme be existentially unforgeable under a chosen message attack~\cite[Ch. 13]{BonehShoup}.

\subsection{Hash Functions and Merkle Trees}

A \textbf{hash function} $\hash$ over $(\mathcal{K}, \mathcal{M}, \mathcal{T})$ is a pair of algorithms $(\hashGen, \hash)$ such that:
\begin{itemize}
    \item $\hashGen(1^{\lambda})$: generates key $s \in \mathcal{K}$; that selects a hash function from the hash family.
    \item $\hash(s, x)$: outputs a string $x' \in {\mathcal{T}}$, which is the evaluation of $\hash(s, \cdot)$ at $x \in \mathcal{M}$.
\end{itemize}
We require that the hash function is collision resistant, meaning it is infeasible to find two different inputs that produce the same hash value~\cite[Ch. 8]{BonehShoup}.
Additionally,  we will use hash functions to construct hash trees, also known as Merkle trees~\cite[Ch. 8]{BonehShoup}.

A Merkle Tree, denoted as $\MT$, consists of three algorithms ($\MTGetRoot, \MTGetProof, \MTVerify$) such that:
\begin{itemize}
    \item $\MTGetRoot(T) \to R$: computes the root $R$ of the tree $T$
    \item $\MTGetProof(T, i) \to \pi_i$: computes the proof $\pi_i$ for the leaf with index $i$
    \item $\MTVerify(T[i], i, \pi_i, R) \to b \in  \{0, 1\}$: verifies the proof $\pi_i$ for the leaf $T[i]$ with index $i$ against the root $R$   
\end{itemize}

\subsection{zk-SNARKs Security Definitions}
\label{app:zksnark}
\newcommand{\advantage}[2]{\textsf{Adv}^{\text{{#1}}}_{{#2}}(\lambda)}
\newcommand{\SIG}{\mathcal{S}}
\newcommand{\pp}{\textsf{pp}}
\newcommand{\rel}{\mathcal{R}}
\newcommand{\NN}{\mathcal{N}}
\newcommand{\verify}{V}
\newcommand{\prover}{P}
\newcommand{\state}{\textsf{st}}
\newcommand{\XX}{\mathcal{X}}
\newcommand{\YY}{\mathcal{Y}}
\newcommand{\abs}[1]{|{#1}|}
\newcommand{\deq}{\mathrel{\mathop:}=}

We define completeness, knowledge soundness, zero-knowledge, non-interactivity, and succinctness for a zk-SNARK below.
\begin{itemize}
\item {\bf Completeness}: if $(x, w) \in \rel$, then verification should pass. That is, for all $\lambda \in \NN$ and all $(x, w) \in \rel$:
\[
            \Pr\left[
                \begin{array}{c}
                \verify(\pp, x,\pi) = 1
                \end{array}
                \ :\ 
                \begin{array}{l}
                     \pp \sample \setup(1^\lambda)\\
                    \pi \gets \prover(\pp, x, w)
                \end{array}
            \right] = 1
\]
\item {\bf Knowledge Soundess}: if an adversary can produce a valid proof for some $x$, then there should be a polytime extractor that can compute a witness $w$ such that $(x, w) \in \rel$. That is, $\Pi$ has knowledge error $\epsilon$ if there exists a PPT extractor $\mathcal{E}$ such that for all PPT $\adv_0, \adv_1$:   
\begin{multline*}
            \Pr\left[
                \begin{array}{c}
                (x, w) \in \rel \\
                \end{array}
                \ :\ 
                \begin{array}{l}
                     \pp \sample \setup(1^{\lambda})\\
                     (x, \state) \sample \adv_0(\pp) \\
                     w \sample \mathcal{E}^{\adv_1(\pp,\state)}(\pp)
                \end{array}
            \right] \geq  \\
            \Pr\left[
                \begin{array}{c}
                \verify(\pp, x, \pi ) = 1 \\
                \end{array}
                \ :\ 
                \begin{array}{l}
                     \pp \sample \setup(1^{\lambda}) \\
                    (x, \state) \sample \adv_0(\pp)  \\
                    \pi \sample \adv_1(\pp, \state)
                \end{array}
            \right] - \epsilon
\end{multline*}
    
\item {\bf Zero-Knowledge}:
We state the definition in the random oracle model where all the algorithms are oracle machine that can query an oracle $H:\XX \to \YY$
for some finite sets $\XX$ and $\YY$.
The zk-SNARK is zero knowledge if there is a PPT simulator $\Pi.\simu$ such that
for all $(x,w) \in \rel$ and all PPT adversaries $\adv$,
the following function is negligible
\[
  \advantage{zk}{\adv,\Pi} \deq \abs{ \begin{aligned}
     \Pr\bigl[\adv^H\bigl(\pp, x, & \prover^H(\pp,x,w)\bigr) = 1\bigr] - \\
   &  \Pr\bigl[\adv^{H[h]}\bigl(\pp, x, \pi)\bigr) = 1\bigr]  
          \end{aligned} }
\]
where $\pp \sample \setup(1^{\lambda})$ and $(\pi,h) \sample \Pi.\simu(\pp,x)$.
Here $h$ is a partial function $h: \XX \to \YY$ output by $\Pi.\simu$,
and $H[h]$ refers to the oracle $H: \XX \to \YY$ modified by entries in $h$.
That is, we allow $\Pi.\simu$ to program the oracle $H$. 

\item {\bf Non-interactive}: 
the proof is non-interactive, and a proof created by the prover can be checked by any verifier.

\item {\bf Succinct}: the proof size and verifier runtime are $o(\abs{w})$.
The verifier can run in linear time in $\abs{x}$. 
\end{itemize}

\section{Private Voting}
\label{app:private_voting}
In the private voting scenario, Alg.~\ref{alg:pr_voting}, the process for a delegate to cast their vote is more nuanced compared to the public vote. The delegate must submit an encrypted vote for each option, a vector $\bm{E}$. The key difference is that for the option they select, they use the encryption of their voting power, while for all other options, they submit an encryption of zeros. To maintain the privacy of their vote, the delegate re-randomizes the encryption of their voting power, as it is public. Additionally, the delegate must provide a proof of correctness for these encrypted votes.

The on-chain contract then verifies the proof and checks if the delegate has not voted previously. After these checks are passed, the contract homomorphically adds the delegate's encrypted votes to the respective tallies for each option. 

The relation used in the proof:
\begin{multline}
\label{eq:rvote}
\rvote := \bigg\{(v, \bm{r}), (\cryptoPK^{\TA}_{\encrypt}, L^{\eid}_d[p_i], \bm{E}, p_i, R_{\eid}, \pi_{p_i}) \mid 
E_v = \encRerand(\cryptoPK^{\TA}_{\encrypt}, L^{\eid}_d[p_i], r_v) \land \\
\forall j \neq v: E_j \gets \enc(\cryptoPK^{\TA}_{\encrypt}, 0; r_j) \land \MTVerify(L^{\eid}_d[p_i], p_i, \pi_{p_i}, R_{\eid}) = 1
\bigg\}
\end{multline}
Table~\ref{tab:rvote} summarizes the witness and public statement for the relation.

\begin{table*}[t]
\makebox[\textwidth][c]{%
\setlength{\tabcolsep}{6pt}
\footnotesize
\begin{tabular}{|l|l|l|l|}
\hline
\multicolumn{4}{|c|}{\textbf{Witness and Public Statement for $\rvote$}} \\ \hline
\multicolumn{2}{|c|}{\textbf{Witness}} & \multicolumn{2}{c|}{\textbf{Public Statement}} \\ \hline
$v \in [3]$ & A vote & $\cryptoPK^{\TA}_{\encrypt} \in \mathbb{G}$ & Encryption public key of TA \\ \hline
$\bm{r} \in \mathbb{Z}_q^3$ & Randomness vector for encryption & $L^{\eid}_d[p_i] \in \mathbb{G}$ & Encrypted voting power of delegate $p_i$ \\ \hline
& & $\bm{E} \in \mathbb{G}^3$ & Encrypted vector of votes \\ \hline
& & $p_i \in [m]$ & Delegate's address and index \\ \hline
& & $R_{\eid} \in \mathbb{F}_p$ & Root of Merkle tree for election $\eid$ \\ \hline
& & $\pi_{p_i} \in \mathbb{F}_p^{\log m}$ & Merkle proof for element at index $p_i$ \\ \hline
\end{tabular}%
}
\caption{Witness and Public Statement for $\rvote$}
\label{tab:rvote}
\end{table*}

\begin{algorithm}[h]
    \caption{Private Voting}
    \label{alg:pr_voting}
    \begin{algorithmic}[1]
    \Function{Voting}{$\eid, L^{\eid}, p_i, v$} \Comment{called by $p_i$}
    \If{$v \in \{\text{yes}, \text{no}, \text{abstain}\}$}
    \State $\bm{r} \sample \mathbb{Z}_q^3$
    \State $E_v \gets \encRerand(\cryptoPK^{\TA}_{\encrypt}, L^{\eid}_d[p_i], r_v)$
    \State $\forall j \neq v: E_j \gets \enc(\cryptoPK^{\TA}_{\encrypt}, 0; r_j)$
    \State $\pi_{p_i} \gets \MTGetProof(L^{\eid}_d, p_i)$
    \State $\pi \gets \proofgen_{\rvote}\left((\cryptoPK^{\TA}_{\encrypt}, L^{\eid}_d[p_i], \bm{E}, p_i, R_{\eid}, \pi_{p_i}), (v, r)\right)$
    \State $\mathsf{vote}[p_i], \bm{E}^{\eid} \gets \Call{\cvoting}{\eid, p_i, \bm{E}, \pi, \pi_{p_i}, L^{\eid}_d[p_i]}$
    \Else
    \State abort
    \EndIf
    \EndFunction 
    \Function{\cvoting}{$\eid$}
    \If {$R_{\eid} \neq 0 \land \dactive[p_i] = 1 \land \mathsf{vote}[p_i] = 0  \land \proofver_{\rvote}\left((\cryptoPK^{\TA}_{\encrypt}, L^{\eid}_d[p_i], \bm{E}, p_i, R_{\eid}, \pi_{p_i}), \pi\right) = 1$}   
    \State $\mathsf{vote}[p_i] = 1$
     \For {$j \in \{\text{yes}, \text{no}, \text{abstain}\}$}
    \State $E^{\eid}_{j} \gets \encAdd( \cryptoPK^{\TA}_{\encrypt}, E^{\eid}_{v}, E_v)$
    \EndFor
    \alglog $\mathsf{vote}[p_i], \bm{E}^{\eid}$
    \Else
    \State abort
    \EndIf
    \EndFunction
    \end{algorithmic}
\end{algorithm}
\section{Security Proof}
\label{app:security}
\subsection{Universal Composability Framework}
For our security analysis, we use a variant of the UC framework called SUC~\cite{C:CanCohLin15}. These frameworks require that no PPT adversary can distinguish between the execution of a real protocol $\pi$ and an execution of an ideal process $f$. 

In every execution, there is an environment~$\mathcal{Z}$, an adversary~$\adv$, a set of parties $p_1, \dots, p_n$, and an ideal process~$f$ (sometimes called an ideal functionality) . The environment $\mathcal{Z}$ writes inputs to parties $p_1, \dots, p_n$, reads their outputs, and can also communicate with the adversary $\adv$. The execution ends once $\mathcal{Z}$ outputs a bit.
This $\mathcal{Z}$ represents all the external protocols that may run concurrently with our protocol. Due to space constraints, we will not describe all aspects of the communication and execution model of SUC, but we will briefly introduce the following special cases of these models.

In the \emph{real model}, there is no ideal functionality, and honest parties adhere to the specified protocol $\pi$. In the \emph{ideal model}, the parties are restricted to only communicate with the ideal functionality~$f$. In the \emph{hybrid model}, both the protocol $\pi$ and ideal functionality~$f$ exist. Honest parties follow the protocol, but in addition, the protocol permits parties to send messages to~$f$ and specifies how to process messages received from $f$.

We say that $\pi$ securely realizes $f$ if, for every ``real-world'' adversary $\adv$ interacting with $\pi$, there exists an ``ideal-world'' adversary $\simu$, such that no environment $\mathcal{Z}$ can distinguish between these two scenarios. A similar statement can be defined for the hybrid model vs. ideal model.

The primary distinction of the SUC framework compared to UC is that SUC incorporates built-in authenticated channels. Additionally, it does not allow the dynamic addition of parties, thus mandating that protocols operate with sets of parties fixed ahead of time. Because of these constraints, the SUC framework cannot accommodate every type of protocol. However, it is compatible with our settings, making it a good choice for our security proof. Furthermore, \cite{C:CanCohLin15} demonstrates a security-preserving transformation from SUC to UC. This essentially implies that our protocol, proven to be SUC-secure, can also be made UC-secure.

\subsection{Security Proof}
We focus on the public voting scenario, and note that our proof can be similarly adapted for the private voting scenario.  In our analysis, we consider a static adversary that can corrupt voters but cannot corrupt the trusted authority~$\TA$. We operate within the local random oracle model, as we require the simulator to program the random oracle~\cite{grom}.
To prove the security of our protocol, we start by defining an ideal process for the voting process, ideal functionality $\fvote$ (defined in Alg.~\ref{alg:ideal}). 
As previously mentioned, we assume the existence of a computing bulletin board, provided by a smart contract. This forces us to use a hybrid model.
We define a contract functionality $\fcontract$ (Alg.~\ref{alg:ercideal}), that captures the computing bulletin board. Operating within the $\fcontract$-hybrid model, we abstract the bulletin board as a functionality that performs only the necessary computations for voting. 
We emphasize that $\fcontract$ encompasses only the public on-chain computations in our protocol, which are reliably executed under the assumption of blockchain security or can be independently verified by any honest participant. Therefore, it is reasonable to model these parts as an ideal functionality.

\begin{algorithm}[h]
    \caption{Functionality $\fvote(\text{aux})$}
    \label{alg:ideal}
    \begin{algorithmic}[1]
    \State Functionality $\fvote(\text{aux}, T)$ runs with voters $p_1, \dots, p_n \in \mathcal{P}$, trusted authority $\TA$ and adversary $\adv$. For every party $p_i \in \mathcal{P}, |\mathcal{P}| = n$,  the functionality maintains a bit $\mathsf{reg}_i \in \{0, 1\}$, integers $d_i \in [n]$ and $t_i \in [B]$. For initialization, set $\mathsf{reg}_i = 0$, $d_i :=i$ for all $i\in [n]$.
    \State -- Upon receiving $(\setup, L_T)$ from $\TA$, check whether $L_T \in [B]^N$ and set $t_i = L_T[i]$ for all $i\in [n]$. Send $(\setup, L_T)$ to all participants.
    \State -- Upon receiving $(\register)$ from $p_i$, set $\mathsf{reg}_i := 1$. Send $(\register, p_i)$ to all participants.
    \State -- Upon receiving $(\unregister)$ from $p_i$, set $\mathsf{reg}_i := 0$. Send $(\unregister, p_i)$ to all participants.
    \State -- Upon receiving $(\delegate, p_j)$ from $p_i$, check whether $\mathsf{reg}_i = 0$. In this case, set $d_i := j$. Send $(\delegate, p_i)$ to all participants.
    \State -- Upon receiving $(\undelegate)$ from $p_i$, set $d_i := i$. Send $(\undelegate, p_i)$ to all participants.
    \State -- Upon receiving $(\esetup, desc, eid)$ from $p_i$, set a bit $\cvote^{eid}_j :=0$ and an integer $t^{eid}_j = 0$ for every $p_j$ and a triple of integers $r^{eid} = (0, 0, 0)$. Store $(eid, p_i)$. Send $(\esetup, desc, eid, p_i)$ to all participants.
    \State -- Upon receiving $(\estart, eid)$ from $p_i$, check whether there is a stored value $(eid, p_i)$. In this case, $\mathbf{for\; all}$ $i \in [n]:$
    \State \hskip3em $\mathbf{if}$ $d_i = j, i \neq j$, $\mathbf{then}$ set  $t^{eid}_j := t^{eid}_j + t_i$
    \State \hskip3em $\mathbf{else}$ set  $t^{eid}_i := t^{eid}_i + t_i\mathbf{\; end \; if}$ 
    \State \hskip1em Send $(\estart, eid, p_i)$ to all participants.
    \State -- Upon receiving $(\cvote, eid, v)$ from $p_i$, check whether $\mathsf{reg}_i = 1$,  $\cvote^{eid}_i = 0$, and  $v \in \{\text{yes}, \text{no}, \text{abstain}\}$. In this case, 
    set $\cvote^{eid}_i := 1$ and $r^{eid}[v] := r^{eid}[v] + t^{eid}_i$. Send $(\cvote, eid, v, p_i)$ to all participants. 
    \State -- Upon receiving $(\tally, eid)$ from $\TA$, 
    for every $v$ set $r^{\eid}_v = 100\frac{r^{\eid}_v}{\sum^3_{v=1} r^{\eid}_v}$ and send $(\tally, eid, r^{eid})$ to all participants. 
    \end{algorithmic}
\end{algorithm}
\begin{algorithm}[h]
    \caption{On Chain Contract Functionality \fcontract}
    \label{alg:ercideal}
    \begin{algorithmic}[1]
     \State Functionality $\fcontract$ runs with voters $p_1, \dots, p_n \in \mathcal{P}$ and trusted authority $\TA$. 
    \State -- Upon receiving $(\setup, \cryptoPK^A_{\encrypt}$, $\cryptoPK^A_{\sig}$, $L_T$, $\sigma_T)$ from $\TA$, execute  \Call{\csetup}{$L_v, \cryptoPK^A_{\encrypt}$, $\cryptoPK^A_{\sig}$, $L_T$, $\sigma_T$}. Send output to all participants.
    \State -- Upon receiving $(\register)$ from $p_i$, execute \Call{\creg}{$p_i$}. Send output to all participants.
    \State -- Upon receiving $(\unregister)$ from $p_i$, execute \Call{\cunreg}{$p_i$}. Send output to all participants.
    \State -- Upon receiving $(\delegate, \bm{ct}, \pi, \pi_{p_i})$ from $p_i$, execute \Call{\cdelegate}{$p_i, \bm{ct}, \pi, \pi_{p_i}$}. Send output to all participants.
    \State -- Upon receiving $(\undelegate, \bm{ct})$, from $p_i$, execute \Call{\cundelegate}{$p_i, \bm{ct}$}. Send output to all participants.
    \State -- Upon receiving $(\esetup, eid, desc)$ from $p_i$, execute \Call{\cesetup}{$p_i,eid, desc$}. Send output to all participants.
    \State -- Upon receiving $(\estart, eid)$ from $p_i$, execute \Call{\cestart}{$p_i,eid$}. Send output to all participants.
    \State -- Upon receiving $(\cvote, eid, p_i, v, \pi_{p_i}, L^{eid}_d[p_i])$ from $p_i$, execute \Call{\cvoting}{$eid, p_i, v, \pi_{p_i}, L^{eid}_d[p_i]$}. Send output to all participants.
    \State -- Upon receiving $(\tally, eid, \res^{eid})$ from $\TA$, execute
    \Call{\ctally}{$eid, \res^{eid}$}. Send output to all participants.
     \end{algorithmic}
\end{algorithm}

Next, we build a simulator, as defined in Alg.~\ref{alg:sim}. The simulator exists within the ideal model.
There, the trusted authority supplies $\fvote$ with the token distribution, $L_T$. The functionality then subsequently broadcasts $L_T$ to all participants. Upon receiving the token distribution from $\fvote$, $\simu$ executes the trusted authority setup as specified by the protocol, acquiring secret keys for encryption and signature schemes. This enables the simulator to sign and decrypt messages on behalf of the trusted authority for its' interactions with $\adv$ (line 2 of Alg.~\ref{alg:sim}).
For every message it receives from the adversary on behalf of a corrupted party, the simulator forwards a corresponding message to the ideal functionality (lines 4-11 of Alg.~\ref{alg:sim}). Possessing the secret key of the authority, it can decrypt the delegation vector $\bm{ct}$ sent by the adversary and relay the delegate's address to the functionality in clear. 
Conversely, for each message received from the ideal functionality, the simulator, on behalf of an honest party,  sends a corresponding message to the adversary (lines 12-20 of Alg.~\ref{alg:sim}). In the delegation step, $\simu$ does not know the delegate's address, only the fact that the voter has delegated. Therefore, to simulate $\bm{ct}$, it encrypts an all-zero vector and simulates the proof of correctness.
For tallying, it takes the result it received from $\fvote$ and simulates a proof of correct decryption.

\begin{algorithm}[h]
    \caption{Simulator $\simu$}
    \label{alg:sim}
    \small
    \begin{algorithmic}[1]
    \State $\simu$ controls the random oracle $\mathcal{O}$, so may assign responses to queries and, therefore, can simulate proofs. For all $i \in |\mathcal{P}| = n$, where $\mathcal{P}$ is a set of all voters, $\simu$ assigns $\bm{ct}_i :=0$.
    Let $\adv$ corrupt a subset $I \in \mathcal{P}$.
    \State -- Upon receiving $(\setup, L_T)$ from $\fvote$, $\simu$ sets $t_i = L_T[i]$ for all $i\in [n]$ and runs \Call{\asetup}{} to obtain $\cryptoPK^{\TA}_{\encrypt}$, $\cryptoPK^{\TA}_{\sig}$, $L_T$, $\sigma_T$ and sends the output to all participants. Additionally, $\simu$ knows the secret keys $\cryptoSK^{\TA}_{\encrypt}$ and $\cryptoSK^{\TA}_{\sig}$. 
    $\simu$ runs \Call{\csetup}{$ \cryptoPK^{\TA}_{\encrypt},\cryptoPK^{\TA}_{\sig}, L_T, \sigma_T$} and sends the output to all participants.
    \State -- $\simu$ invokes $\adv$ and simulates $\fcontract$ on valid calls, ignores all invalid calls. 
    \State -- \textbf{For every $p_i \in I$:}
    \statetab --  Upon receiving $(\register)$ from $p_i$,  $\simu$ sends $(\register)$ on behalf of $p_i$ to $\fvote$
    \statetab --  Upon receiving $(\unregister)$ from $p_i$,  $\simu$ sends $(\unregister)$ on behalf of $p_i$ to $\fvote$
    \statetab --  Upon receiving $(\delegate, \bm{ct}, \pi, \pi_{p_i})$ from $p_i$,  $\simu$ checks whether the proofs $\pi, \pi_{p_i}$ are valid. If  $\pi_{p_i}$ is a valid proof but the leaf opening is not $t_i$, $\simu$ aborts. Otherwise, it decrypts $\bm{ct}$. $\simu$ sets $\bm{ct}_i = \bm{ct}$ and sends $(\delegate, p_j)$ on behalf of $p_i$ to $\fvote$, where $p_j$ is such that $\bm{ct}[p_j] \neq 0$. 
    \statetab --  Upon receiving $(\undelegate, \bm{ct})$ from $p_i$. If $\bm{ct}_i \neq \bm{ct}$ but $\hash(\bm{ct}_i) = \hash(\bm{ct})$, then $\simu$ aborts. If $\bm{ct}_i = \bm{ct}$, it sends $(\undelegate)$ on behalf of $p_i$ to $\fvote$.
    \statetab -- Upon receiving $(\esetup, eid, desc)$ from $p_i$,  $\simu$ sends $(\esetup, desc, eid)$ on behalf of $p_i$ to $\fvote$
    \statetab -- Upon receiving $(\estart, eid)$ from $p_i$, $\simu$ sends $(\estart, eid)$ on behalf of $p_i$ to $\fvote$
    \statetab -- Upon receiving $(\cvote, eid, p_i, v, \pi_{p_i}, L^{eid}_d[p_i])$ from $p_i$,  $\simu$ checks whether the proofs $\pi$ and $\pi_{p_i}$ are valid.  If $\pi_{p_i}$ is a valid proof but the leaf opening is not $L^{eid}_d[p_i]$, $\simu$ aborts. Otherwise, $\simu$ sends $(\cvote, eid, v)$ on behalf of $p_i$ to $\fvote$.
    \State -- \textbf{For every $p_i \in \mathcal{P} \setminus I$:}
    \statetab --  Upon receiving $(\register, p_i)$ from $\fvote$, $\simu$ executes \Call{\creg}{$p_i$} and sends the output to $\adv$.
    \statetab --  Upon receiving $(\unregister, p_i)$ from \fvote, $\simu$ executes \Call{\cunreg}{$p_i$} and sends the output to $\adv$.
    \statetab --  Upon receiving $(\delegate, p_i)$ from \fvote, $\simu$  produces $\bm{ct}$ as an encryption of all-zero vector, generates a Merkle proof $\pi_{p_i}$ and simulates a zk-proof $\pi.$ Then, $\simu$ executes \Call{\cdelegate}{$p_i, \bm{ct}, \pi, \pi_{p_i}$} and sends the output to $\adv$.
    \statetab --  Upon receiving $(\undelegate, p_i)$ from \fvote, $\simu$ executes \Call{\cundelegate}{$p_i, \bm{ct}$} and sends the output to $\adv$.
    \statetab -- Upon receiving $(\esetup, desc, eid, p_i)$ from \fvote,  $\simu$ executes \Call{\cesetup}{$p_i, eid, desc$} and sends the output to $\adv$.
    \statetab -- Upon receiving $(\estart, eid, p_i)$ from $\fvote$, $\simu$ executes \Call{\cestart}{$p_i, eid$} and sends the output to $\adv$.
    \statetab -- Upon receiving $(\cvote, eid, p_i, v, p_i)$ from $\fvote$,  $\simu$ generates a Merkle proof $\pi_{p_i}$, executes \Call{\cvoting}{$eid, p_i, v, \pi_{p_i}, L^{eid}_d[p_i]$} and sends the output to $\adv$.
    \statetab -- Upon receiving $(\tally, eid, r^{eid})$ from $\fvote$, $\simu$ simulates a zk-proof $\pi$. Then, $S$ executes \Call{\ctally}{$eid, \res^{eid}, \pi$} and sends the output to $\adv$.
    \end{algorithmic}
\end{algorithm}

Finally, we are ready to prove our main security theorem.
\begin{theorem}
$\Kite$ (Sec.~\ref{sec:protocol})
SUC-securely realizes $\fvote$ with respect to $\Pi^{\fcontract}$, 
assuming a collision-resistant hash function, 
a CPA-secure encryption scheme, and 
a secure non-interactive zero-knowledge argument of knowledge.
\end{theorem}

\begin{proof}
For our analysis, we apply the hybrid argument technique. The goal is to start with our protocol and introduce modifications, step by step, gradually transforming into the ideal functionality. Note that SUC assumes the authenticated channels, therefore, we do not need to verify trusted authority's signature explicitly, as it is implicitly verified in the model.

\textbf{Experiment~0.}
Experiment~0 is the same as the protocol, with the exception that we restrict the adversary $\adv$ to generate Merkle proofs honestly.

$\sucHybrid^{\fcontract}_{\Pi, \adv, \mathcal{Z}} \approx \expr_0$. 
We argue that the protocol and Experiment~0 are indistinguishable to $\mathcal{Z}$. To show that, assume the opposite, so the protocol and Experiment~0 are not indistinguishable to $\mathcal{Z}$. The only difference is the ability of $\adv$ to generate valid Merkle proofs for invalid leaf values in $\sucHybrid^{\fcontract}_{\Pi, \adv, \mathcal{Z}}$.
Therefore, we can build a new adversary $\mathcal{B}$, that uses $\mathcal{Z}$ to find $p_i, p'_i$ such that $p_i \neq p'_i$ but $\hash(p_i) = \hash(p'_i)$.
However, due to the collision-resistance property of the underlying hash function, this can happen with a negligible probability only.

\textbf{Experiment~1.}
Experiment~1 is the same as Experiment~0, except that we restrict the adversary $\adv$ further and require it to send the correct $\bm{ct}$ in the undelegation step.

$\expr_0 \approx \expr_1$. 
We argue that the Experiment~0 and Experiment~1 are indistinguishable to $\mathcal{Z}$. To show that, assume the opposite, so the Experiment~0 and Experiment~1 are not indistinguishable to $\mathcal{Z}$. The only difference is the ability of $\adv$ to find a $\bm{ct}$ such that $\bm{ct} \neq \bm{ct}_i$ but $\hash(\bm{ct}_i) = \hash(\bm{ct})$. Again, we can build a new adversary $\mathcal{B}$, that uses $\mathcal{Z}$ to find $\bm{ct}, \bm{ct}_i$ such that $\bm{ct} \neq \bm{ct}_i$ but $\hash(\bm{ct}_i) = \hash(\bm{ct})$. 
Due to the collision-resistance property of the hash function, this can happen with a negligible probability only.





\textbf{Experiment~2.}
Experiment~2 is the same as Experiment~1, except the following change in \textproc{Delegation} -- the proof $\pi$ is generated by $\mathcal{S}$.

$\expr_2 \approx \expr_1$. 
Experiment~2 and Experiment~1 are indistinguishable to $\mathcal{Z}$.
To show that, assume the opposite, so the Experiment~2 and Experiment~1 are not indistinguishable to $\mathcal{Z}$. Therefore, we can build an adversary $\mathcal{B}$, that uses $\mathcal{Z}$ to distinguish
between $\pi$ simulated by $\mathcal{S}$ and $\pi$ generated by $p_i$.
Due to the zero-knowledge property of the proof system, this can happen with a negligible probability only.

\textbf{Experiment~3}. Is the same as Experiment~2, except the following change in \textproc{Delegation} -- instead of generating $\bm{ct}$ as described in Alg.~\ref{alg:delegation}, set $\bm{ct} :=\enc(\cryptoPK^{\TA}_{\encrypt}, \bm{0}; \bm{r})$, where $\bm{r}$ is drawn randomly.

$\expr_3 \approx \expr_2$. We argue that the Experiment~3 and Experiment~2 are indistinguishable to $\mathcal{Z}$. To show that, assume the opposite, so the Experiment~3 and Experiment~2 are not indistinguishable to $\mathcal{Z}$. Therefore, we can build an adversary $\mathcal{B}$, that uses $\mathcal{Z}$ to distinguish
between $\enc(\cryptoPK^{\TA}_{\encrypt}, \bm{0}; \bm{r})$ and $\enc(\cryptoPK^{\TA}_{\encrypt}, \bm{t}; \bm{r'})$.
Due to the CPA security of the encryption scheme, this can happen with a negligible probability only.

$\expr_3 \approx \sucIdeal_{\fvote, \simu, \mathcal{Z}}$. The Experiment~3 is the simulated interaction of $\mathcal{Z}$ and $\simu$. 
Note that all zero-knowledge proofs in the Experiment 3 are simulated, and the delegated encryption is generated as if encrypting an all-zero vector, exactly like $\simu$ does.
Importantly, $\simu$ never aborts, as we restricted $\adv$ in the Experiments 0 and 1. 
\end{proof}

\subsection{Implementation and Evaluation}
\label{app:implementation}
\subsection{System Implementation Details}
We developed a proof-of-concept implementation of $\Kite$,
which supports private delegation and public voting.  Our implementation totalled 10,238 lines of code and includes a React frontend (2,450 lines), two servers written in Rust (3,312 lines), zero-knowledge circuits written in Noir (911 lines), and on-chain Ethereum smart contracts developed in Solidity (4,476 lines). 
The frontend was designed to mimic the Nouns DAO tally UI as a concrete use case.
The two Rust servers manage all cryptographic operations and communications with the chain. 
The first Rust server functions as the backend tasked with constructing the zero knowledge proofs as well as deploying and calling most on-chain contracts with the relevant data required for verification. 
The second Rust server represents the trusted authority~$\TA$, responsible for setting up the on-chain private governance contract and decrypting the final tallies. 
In practice, the trusted authority would be securely distributed among multiple parties using threshold decryption. However, for our proof-of-concept, the trusted authority is implemented as a single entity. 

Our on-chain Solidity implementation builds on the Open Zeppelin ERC20 and Compound Governor Bravo contracts.
We extended the ERC20 contract to include a locking mechanism, to restrict the movement of tokens when needed. We utilize a mapping from addresses to booleans, indicating the lock status for each user. The standard ERC20 functions -- transfer, transferFrom, approve, and spendAllowance -- have been modified to revert if the user's tokens are locked, while retaining their original functionality otherwise. We use COMP tokens as the ERC20 governance tokens. 
To vote on proposals, every voter must delegate their voting power to themselves or to some other voter (the delegate). 
When a proposal is first posted, the Comp contract (which implements the COMP token) takes a snapshot of the current voting powers and delegation status of all voters.
When a voter casts their vote, the Comp 
contract sends the snapshot of the user's voting power, as recorded when the proposal was first posted, to the Governor Bravo contract.

The voting protocol logic is implemented as a significant 
extension to the Governor Bravo contract. This includes the logic
to verify all the ZK proofs generated by the Rust servers
on behalf of the participants, as well as the logic to act homomorphically
on encrypted data. 
The original Governor Bravo contract permitted users to delegate their voting power while simultaneously receiving delegations from others. Our delegate registration implementation eliminates this potential loophole. 
Additionally, when a user who previously delegated thier voting power wishes to redelegate to a new account, there was no explicit undelegation step; instead, users directly delegated to another individual. Our implementation rectifies this by incorporating an undelegation step, enabling users to subtract their token balance from their original delegate.

To reduce the time to generate the ZK proofs, our implementation introduces the notion of an \emph{anonymity set}.  
When a voter intends to delegate, the backend server constructs an anonymity set
by randomly sampling $T-1$ delegate accounts without replacement, and adding the delegate's account to obtain an anonymity set of size~$T$, where $T$ is chosen by the voter  The voter's posted vector of encrypted powers is now of size $T$, which is smaller than
the total number of delegates in the system.  This both reduces the amount of data to post on chain, and reduces the time to generate the relevant ZK proofs. 
The cost is that an observer learns that the voter delegated to someone in the anonymity set, whereas in the full protocol of Section~\ref{sec:protocol}, an observer only learns that a delegation to some delegate took place. Note that if the voter chooses $T=1$, then it essentially become public delegation.

On election start, the off-chain, trusted authority~$\TA$ generates the election snapshot and pushes it onto the blockchain. The snapshot is stored on-chain and written to the transaction log to facilitate off-chain access to the snapshotted voting powers.

In the tally decryption step, the trusted authority~$\TA$ decrypts the tally for a specific election and computes the percentages of votes for, against, and abstained. These percentages, along with a zkproof of correct decryption are submitted on-chain. Posting percentages instead of raw or rounded vote weights minimizes the disclosure of information regarding a delegate's voting power, while still ensuring transparency in the tally decryption process.

\ignore{
To showcase all of the algorithms in our governance protocol, we crafted a setup script. This script initializes the master verifier and the ERC-20 Private token. After initializing the Private Governance contract, we execute a backend function that registers 20 accounts as delegates, each endowed with Private token balances. This mass delegate registration step is crucial for establishing real accounts that contribute to the anonymity set for delegation. Following this setup, users gain the ability to interact seamlessly with our entire protocol, leveraging an anonymity set of any desired size.
}

\subsection{Implementing the ZK Relations}

We implemented each of the ZK relations used in Section~\ref{sec:protocol}, including $\rvote$ (used for private voting), though we only support public voting in our proof-of-concept.   While in Section~\ref{sec:protocol} we assumed a fully trusted authority~$\TA$ to simplify the proof of security, our implementation relaxes this assumption somewhat.  In particular, we require the authority to provide a ZK proof of correct decryption of the final tally results.  We do so using the following $\rdec$ relation.
\begin{align*}
\rdec := \bigg\{\left(\cryptoSK^{\TA}_{\encrypt},\ (\cryptoPK^{\TA}_{\encrypt}, \bm{E}^{\eid}, \res^{\eid})\right) \quad \mid \quad \text{for $i \in [3]$: } 
D^{\eid}_i \gets \dec(\cryptoSK^{\TA}_{\encrypt}, E^{\eid}_i) \land \\  \res^{\eid}_i = 100\frac{D^{\eid}_i}{\sum^3_{i=1} D^{\eid}_i}, \ \ \bigg\}
\end{align*}
where $\cryptoSK^{\TA}_{\encrypt} \in \mathbb{Z}_q$ and $\cryptoPK^{TA}_{\encrypt} \in \mathbb{G}$ are the secret and public keys of~$\TA$, 
$\bm{E}^{\eid} \in \mathbb{G}^3$ is a vector of the encrypted numbers of votes for each option,
and $\res^{\eid} \in [0, 100]$  is a vector of the precentage of votes for each option.

\ignore{
Table below
summarizes the witness and public statement for $\rdec$.
\begin{table}[h]
\footnotesize
\label{table:rdec}
\begin{tabular}{|l|p{6.5cm}|}
\hline
\multicolumn{2}{|c|}{\textbf{Witness}} \\ \hline
$\cryptoSK^{\TA}_{\encrypt} \in \mathbb{Z}_q$ & The encryption secret key of the trusted authority \\ \hline
\multicolumn{2}{|c|}{\textbf{Public Statement}} \\ \hline
$\cryptoPK^{TAa}_{\encrypt} \in \mathbb{G}$ & The encryption public key of the trusted authority \\ \hline
$\bm{E}^{\eid} \in \mathbb{G}^3$ & A vector of the encrypted numbers of votes for each option \\ \hline
$\res^{\eid} \in [0, 100]$  & A vector of the precentage of votes for each option \\ \hline
\end{tabular}
\caption{Witness and Public Statement for $\rdec$}
\end{table}
}

We implemented all zero-knowledge relations   using the Noir language~\cite{noir}. Noir is a domain-specific language with support for a modular backend meant to work with any ACIR (Abstract Circuit Intermediate Representation)-compatible proving system. We use Aztec Labs' Barretenberg backend for our proving system, which runs on PLONK \cite{plonk}. Noir is a powerful tool for rapid code iteration in circuits that deal with complex operations, while still providing reasonable performance. This makes Noir well-suited for our needs. One limitation of Noir is its lack of support for elliptic curve operations on many familiar curves such as secp256k1. Thus our implementation is centered around the use of the BabyJubJub curve~\cite{BabyJubJub}. 
The restriction to arithmetic on this curve necessitates the use of a few extra proofs in the implementation in order to manage gas costs, since these curve operations are not currently implemented in Solidity efficiently and robustly.
Specifically, we make use of the following relations to offload on-chain work to our backend rust server:

\begin{align*}
&\mathcal{R}_{\text{addmt}} := \{(\emptyset), (\cryptoPK^{\TA}_{\encrypt}, ct_1, ct_2, ct_{+}, \pi_{p_i}, R)
\mid
\MTVerify(ct_2, p_i, \pi_{p_i}, R) = 1 \land ct_{+} =  \encAdd( \cryptoPK^{\TA}_{\encrypt}, ct_1, ct_2) \} \\
&\mathcal{R}_{\text{vecsub}} := \{(\emptyset), (\cryptoPK^{\TA}_{\encrypt}, \bm{ct_1}, \bm{ct_2}, \bm{ct_{-}})
\mid 
\bm{ct_{-}} =  \encAdd( \cryptoPK^{\TA}_{\encrypt}, \bm{ct_1}, -\bm{ct_2})
)
\}\\
&\mathcal{R}_{\text{vecadd}} := \{(\emptyset), (\cryptoPK^{\TA}_{\encrypt}, \bm{ct_1}, \bm{ct_2}, \bm{ct_{+}})
\mid 
\bm{ct_{-}} =  \encAdd( \cryptoPK^{\TA}_{\encrypt}, \bm{ct_1}, \bm{ct_2})
)
\}\\
&\rencsub := \{(\emptyset), (\cryptoPK^{\TA}_{\encrypt}, t, ct, ct_{-})
\mid 
e = \enc(\cryptoPK^{\TA}_{\encrypt}, -t; 0) \land ct_{-} =  \encAdd( \cryptoPK^{\TA}_{\encrypt}, ct, e) \}\\
&\renc = \{(\emptyset), (\cryptoPK^{\TA}_{\encrypt}, ct, t, r)
\mid
ct =  \enc(\cryptoPK^{\TA}_{\encrypt}, t; r)
\}
\end{align*}

In particular, $\mathcal{R}_{\text{addmt}}$ is used by Alg.~\ref{alg:voting}, $\mathcal{R}_{\text{vecsub}}$ by Alg.~\ref{alg:delegation}, $\rencsub$ and $\renc$ by Alg.~\ref{alg:dreg}. As a special case, Alg.~\ref{alg:delegation} requires the homomorphic addition of ciphertext vectors (for which we would verify a proof of $\mathcal{R}_{\text{vecadd}}$) immediately after verifying the relation $\rdel$. In our implementation, we actually concatenate $\rdel$ and $\mathcal{R}_{\text{vecadd}}$ into one circuit to avoid the consecutive execution of two expensive proof verifications on-chain. However, since $\rdel$ is inherent to the protocol and $\mathcal{R}_{\text{vecadd}}$ is implementation-specific, we report their metrics separately below. These proofs allow us to carry out expensive operations lacking robust implementations in Solidity within our Rust backend. The results of these operations, as well as a proof of their correctness, are then provided to the on-chain contract.

All of the relations we use must be verified on-chain. Noir's tooling allows for the generation of smart contracts with functionality to load a relation-specific verification key and verify a set of public inputs against a proof, with each contract corresponding to a single relation. We extend these auto-generated contracts by rolling them into a single contract (3,291 lines of code, most of which are generated by Noir) that has the ability to load one of many verification keys and verify any of the relations we use in our implementation. This reduces the number of helper contracts referenced by our governance contract, which in turn reduces gas costs and eliminates duplicated code across contracts.
As part of the delegation process, a user submits a public list of delegate addresses. An observer knows that the user delegated their voting power to one of these addresses, but not which one in particular. Thus, in $\Kite$, the relation $\rdel$ depends on the size of the anonymity set, defined in Section~\ref{sec:implementation}.
For the sake of simplicity in our proof-of-concept implementation, we provide users the option to use an anonymity set of size 5, 10, or 20. However, adding support for further set sizes is trivial. Smaller anonymity sets significantly reduce proving time and gas costs but offer less privacy. Conversely, larger sets provide greater privacy at increased computational cost, leaving users to balance these trade-offs based on their privacy needs.

The time cost of the implementation is important to the usability of the protocol. Since the verification time is constant, we took steps to reduce the proving time of the relations. We use the zk-friendly Poseidon hash function \cite{AFRICACRYPT:GraKhoSch23} in all necessary Merkle proofs, and implement functions like ElGamal encryption within Noir circuits ourselves to take advantage of speedups afforded by details of our protocol. For example, many scalar-point multiplies can be foregone when the randomness is deterministic, as is the case in Alg.~\ref{alg:dreg}. We also implement a custom "small scalar"-point multiplication using the Double-and-Add algorithm. This removes unnecessary loop iterations, as the Noir interface for this operation only supports 254 bit scalars while we often only need 32 bit.

\subsection{Evaluation Details}
\label{app:evaluation}
\begin{figure*}[t!]
    \centering
    \begin{subfigure}[t]{0.5\textwidth}
        \centering
        \includegraphics[scale=0.27]{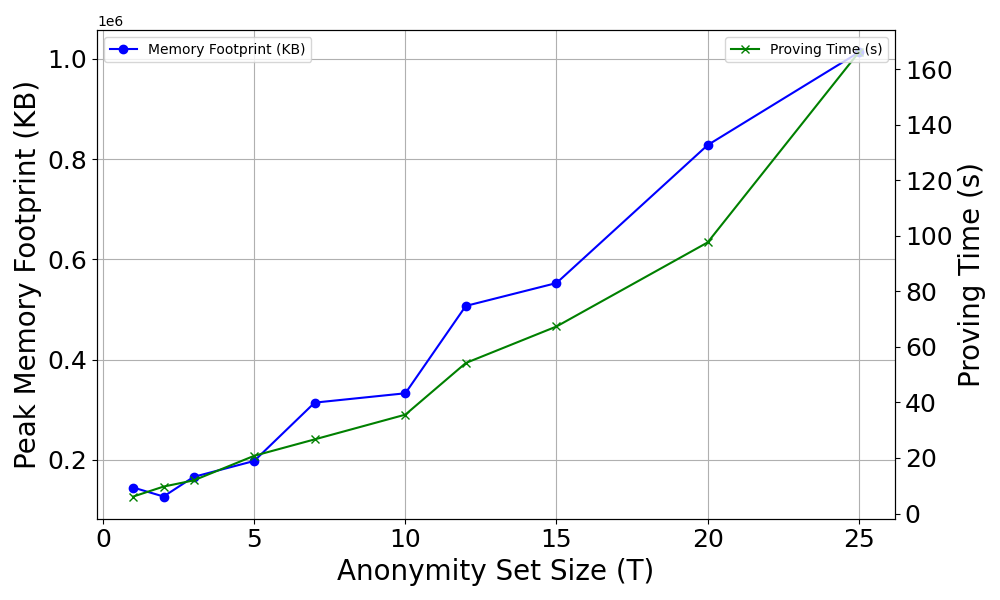}
        \caption{Proving time of $\rdel$ and peak memory footprint of proof generation vs anonymity set size.}
        \label{fig:anonymity}
    \end{subfigure}%
    \begin{subfigure}[t]{0.5\textwidth}
        \centering
        \includegraphics[scale=0.269]{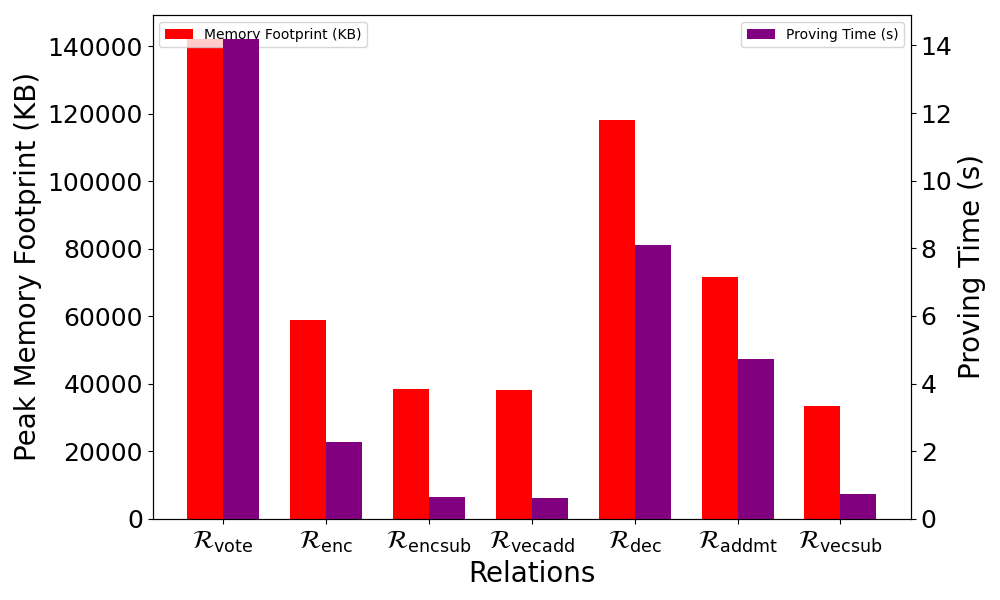}
        \caption{Proving times and peak memory footprints of proof generation of all relations, except $\rdel$.}
        \label{fig:relations}
    \end{subfigure}
    \caption{Proving time and peak memory footprint of proof generation.}
    \label{fig:combined}
\end{figure*}

Our performance figures were collected on a consumer-grade device (MacBook Air) running an 8-core Apple M2 chip with 8GB of RAM to emulate the conditions end-users of the protocol may observe. We provide proving time and peak memory footprint data for each of the relations in our proof-of-concept implementation (Figures~\ref{fig:anonymity} and~\ref{fig:relations}), as well as gas cost data for each step of the protocol in ~\cref{sec:implementation}.

All proving times are computed as an average of ten runs using the Nargo CLI provided by Aztec Labs. For $\rdel$, we report the metrics as a function of the size of the anonymity set, and collected results for sets of size up to 25. As mentioned previously, in the full implementation we only make use of the circuits for sizes 5, 10, and 20. We see that the proving time and associated memory cost for $\rdel$ increases with the anonymity set, as expected. The largest anonymity set size we tested took just over 2 minutes and 46 seconds to prove. However, sets of intermediate sizes 15 and 12 were much faster, with proving times slightly over and below 1 minute, respectively (Fig.~\ref{fig:anonymity}). We observed that the inclusion of a small-scalar multiplication implementation reduces the proving time of $\rdel$ by an average of 42.61\% across all anonymity set sizes. The results for $\rdel$ also provide insight into the cost of private voting, as casting a vote with a small number of options incurs roughly similar or lower proving costs than generating a delegation vector of that size.
Out of the other relations, $\rvote$ is the most costly, with the additional implementation-specific and protocol-independent relations $\mathcal{R}_{\text{addmt}}, \mathcal{R}_{\text{vecsub}}, \mathcal{R}_{\text{vecadd}}, \mathcal{R}_{\text{encsub}}$, and $\mathcal{R}_{\text{enc}}$ claiming only marginal resources in comparison (Fig.~\ref{fig:relations}). The proving times of the vector-valued $\mathcal{R}_{\text{vecsub}}$ and $\mathcal{R}_{\text{vecadd}}$ would depend on the size of the anonymity set, but in our implementation we use the circuits for a vector of length 20 and pad to fill the vector for smaller set sizes since these relations are quite lightweight.

We also examine the circuit size, and find that the order of relations by proving time is the same as their circuit size. The largest circuit we implemented was $\rdel$ with an anonymity set size of 25, with a reported 566,597 gates (767,297 without small-scalar multiplication); the smallest being the implementation-specific 20-element $\mathcal{R}_{\text{vecadd}}$ with 1,325 gates.

The proofs posted on-chain are a constant 4,288 bytes in size. Gas costs to verify the different relations averaged to 406,646 with a median of 396,323 on Forge's Anvil local testnet. Gas cost optimization was not a primary focus of this work, leaving room for reduction in future work. In our implementation, multiple relations are concatenated into a single larger relation, allowing one proof to verify multiple operations, such as $\rdel$ and $\mathcal{R}_{\text{vecsub}}$, reducing gas costs.
Since we also used additional proofs to off-load the elliptic curve operations to the Rust backend, implementing operations on the chosen curve efficiently in Solidity to avoid the use of additional proofs can result in future gas savings. However, the use of a single master verifier contract did significantly reduce the gas costs associated with the deployment of verifier contracts in our implementation by eliminating duplicated information on the blockchain. If the verifier contracts for each of the eight circuits we used were deployed seperately as they are generated by Noir's tooling, it would cost an estimated 18,567,088 gas (according to estimates from Forge gas reports \cite{foundry}). This is compared with our master verifier's deployment cost of 5,981,566 gas, saving 12,585,522 gas during contract deployment. 

Our proof-of-concept achieves reasonable performance on a consumer-grade machine with minimal optimization. Most proofs are generated within 15 seconds, except delegation, which takes 7–167 seconds based on the desired privacy levels. However, delegation is rather infrequent, as its main purpose is to reduce user interaction with the voting system, leaving end-users to primarily engage with lower-latency operations.

\end{document}